\documentclass[twocolumn]{aastex631}

\usepackage{amsmath}
\usepackage{amsfonts}
\usepackage{graphicx}
\usepackage{hyperref}
\usepackage{float}
\usepackage{multirow}
\usepackage[mathlines]{lineno}

\begin{document}

\title{Distinguishing AGN Feedback Models with the Thermal Sunyaev-Zel'dovich Effect }

\author{Skylar Grayson}
\affiliation{School of Earth and Space Exploration, Arizona State University, P.O. Box 876004, Tempe, AZ 85287, USA}

\author{Evan Scannapieco}
\affiliation{School of Earth and Space Exploration, Arizona State University, P.O. Box 876004, Tempe, AZ 85287, USA}

\author{Romeel Dav\'e}
\affiliation{Institute for Astronomy, Royal Observatory, University of Edinburgh, Edinburgh EH9 3HJ, UK 4}
\affiliation{Department of Physics and Astronomy, University of the Western Cape, Bellville, Cape Town 7535, South Africa}
\affiliation{South African Astronomical Observatories, Observatory, Cape Town 7925, South Africa}

\accepted{September 24, 2023}
\submitjournal{The Astrophysical Journal}

\begin{abstract}

Current models of galaxy formation require strong feedback from active galactic nuclei (AGN) to explain the observed lack of star formation in massive galaxies since $z \approx 2,$ but direct evidence of this energy input is limited. We use the SIMBA cosmological galaxy formation simulations to assess the ability of thermal Sunyaev-Zel'dovich (tSZ) measurements to provide such evidence, by mapping the pressure structure of the circumgalactic medium around massive $z \approx 0.2- 1.5$ galaxies.  We undertake a stacking approach to calculate the total tSZ signal and its radial profile in simulations with varying assumptions of AGN feedback, and we assess its observability with current and future telescopes.  By convolving our predictions with the 2.1' beam of the Atacama Cosmology Telescope (ACT), we show that current observations at $z \approx 1$ are consistent with SIMBA's fiducial treatment of AGN feedback, and inconsistent with SIMBA models without feedback. At $z \approx 0.5$, observational signals lie between SIMBA run with and without AGN feedback, suggesting AGN in SIMBA may inject too much energy at late times. By convolving our data with a 9.5'' beam corresponding to the TolTEC camera on the Large Millimeter Telescope Alfonso Serrano (LMT), we predict a unique profile for AGN feedback that can be distinguished with future higher-resolution measurements. Finally, we explore a novel approach to quantify the non-spherically symmetric features surrounding our galaxies by plotting radial profiles representing the component of the stack with m-fold symmetry.

\end{abstract}


\section{Introduction}
\label{sec:intro}

A major mystery of modern extragalactic astronomy is the so-called cosmic downsizing that occurred from $z \approx 2$ to today \citep{cowie, drory}. In this period, the characteristic mass of star-forming galaxies decreased by a factor of roughly ten, and the characteristic luminosity of active galactic nuclei (AGN) decreased by a factor of roughly 1,000 \citep{Treu_2005, drory, hopkins}. This goes against expectations from hierarchical models of galaxy formation, which predict that accretion and merging led to ever-increasing masses of star-forming galaxies and active black holes over time \citep{rees, whitefrenk}.  

As downsizing is observed in both the characteristic stellar masses and AGN luminosities, it is postulated that there is a single mechanism leading to the observed evolution of both these populations \citep[e.g.][]{2012feedback}.  The leading model for this mechanism is feedback from AGN, which ejects energy and material into the surrounding circumgalactic medium at a level that is likely to be able to suppress gas accretion \citep{Scannapieco_2004,2005Natur.433..604D, 2006MNRAS.365...11C, 2006ApJ...653...86T,dubois,2015MNRAS.446..521S, 2018MNRAS.473.4077P, dave}.  However,  the details of this process remain largely unknown, due both to the large range of scales involved and the lack of understanding of the key physical processes \citep{obs_agn}. 

Nevertheless, all modern galaxy evolution simulations rely on some form of AGN feedback to reproduce observations. These include Horizon-AGN \citep{horizon}, EAGLE \citep{eagle}, IllustrisTNG \citep{illustris}, and SIMBA \citep{dave}, which all have a different treatment of material ejection and initial conditions. As direct measurements of AGN energy input are uncertain \citep{2009ApJ...706..525M,2011MNRAS.410.1957H,2013MNRAS.436.3286A,2013ApJ...762...49B,2019NatAs...3..265H}, a promising alternative approach is observing the gas affected by feedback. 

One method for probing this gas is through the measurement of the imprint of the circumgalactic medium (CGM)  on small-scale  Cosmic Microwave Background (CMB) fluctuations.  While Silk damping washes out primary anisotropies from early times at scales below 10 arcmin, interactions with structures that developed since recombination impose a variety of secondary anisotropies that can be used to understand later-time structure formation. 

One important signature occurs when photons pass through hot and dense gaseous regions, and inverse Compton scattering increases the energy of CMB photons. This  is known as the thermal Sunyaev-Zel'dovich (tSZ) effect, and its impact  on the temperature of the CMB has a distinctive redshift-independent frequency dependence
\begin{equation} \label{overview_tsz}
    \frac{\Delta T}{T_{\text{CMB}}}= y \left( x \frac{e^x+1}{e^x-1}-4  \right),
\end{equation}
where the Compton-$y$ parameter is given by 
\begin{equation}
\label{eq1}
    y = \sigma_T \int dl \ n_e \frac{k(T_e-T_{\rm CMB})}{m_e c^2},
\end{equation}
and $\sigma_T$ is the Thomson cross section,  $n_e$ is the electron number density, $k$ is the Boltzmann constant, $T_e$ is the electron temperature, $T_{\rm CMB}=2.725$K is the CMB temperature, $m_e$ is the electron mass, $c$ is the speed of light, and $x = \frac{h \nu}{k T_{\rm CMB}} = \frac{\nu}{56.81 \text{ GHz}}$ with $h=$ Planck's constant.
A useful feature of the tSZ signal is it provides a measure of the line-of-sight integral of the pressure, which allows a calculation of the thermal energy in the CGM \citep{Scannapieco_2008,Spacek_2017,sz_review}, as
\begin{equation}
\label{eq:therme}
    E_{\text{therm}} = 2.9 \times 10^{60} \left( \frac{l_{\text{ang}}}{\text{Gpc}}\right)^2 \frac{\int \Delta y (\theta) \, \theta \, d\theta}{10^{-6} \text{arcmin}^{-2}},
\end{equation}
where $l_{\text{ang}}$ is the angular diameter distance and $\int \Delta y (\theta) d\theta$ is the total integrated $y$ signal in the region of interest. 

While there has been extensive research into observing the tSZ effect around galaxy clusters \citep[e.g.][]{planck_tsz, planck_tsz_2016, Hilton_2018,  Lokken_2022}, studying lower-mass halos has proven more difficult. Some quasars are bright enough for individual detections \citep[e.g.][]{lacy_2018, abdulla_2019, brownson_2019}, but a stacking approach is needed to extract the tSZ signal for most galaxy samples. \cite{hall_2019} used a combination of Atacama Cosmology Telescope (ACT), Herschel Space Observatory, and Very Large Array data to stack on a sample of quasars, measuring the tSZ effect with a significance of $3.8\sigma$ at z$>$ 1.91. Nearby massive  galaxies (z $\lesssim$ 0.35) were stacked in \cite{Greco_2015}, detecting a signal for galaxies more massive than $10^{11.1-11.3} M_\odot$. More recently, \cite{Schaan_2021} stacked on galaxy groups at redshifts from 0.3 to 0.55, finding a $10-12.9 \sigma$ detection. 

At higher redshift, a stacking analysis of $z\approx 1$ quiescent galaxies was conducted with South Pole Telescope (SPT) data in \cite{Spacek_2016} and \cite{Spacek_2017}, finding a $\approx$ 2-3 $\sigma$ signal indicating non-gravitational heating. More recently, \cite{jeremy} used Wide-Field Infrared Survey Explorer (WISE), the Dark Energy Survey, and SPT data to build off the results, finding a $10.1\sigma$ signal around $z\approx 1$ galaxies. The same group built on their analysis with additional data from ACT, detecting tSZ signals at $10.3\sigma$ \citep{meinke_2023}.

There has also been extensive computational work modeling the tSZ effect using cosmological simulations. Early work looked at structure as a function of angular scale through the tSZ power spectrum \citep{Hobson_1996, 2000PhRvD..61l3001R, 2000MNRAS.317...37D,Springel_2001, Zhang_2002,Roncarelli_2007}, but more recently the tSZ signal has been constructed by identifying regions around galaxies and halos constraining feedback \citep{Scannapieco_2008}. 

\cite{Spacek_2018} compared SPT and ACT results against the Horizon-AGN and Horizon-NoAGN simulations, finding that Horizon-NoAGN provided a better fit to both ACT and SPT measurements than Horizon-AGN, while lower-redshift samples were more closely aligned than those at high redshifts. \cite{Kukstas_2020} used EAGLE and BAHAMAS in comparison with Planck tSZ maps and found that the quenching treatment in those simulations may be too efficient. \cite{moser_2022} compared results from the IllustrisTNG- and SIMBA-based Cosmology and Astrophysics with MachinE Learning Simulations (CAMELS) simulations against ACT measurements at $z \approx 0.5$, finding good agreement at radii $<$ 2', but at discrepancy between the observations and simulations at larger radii.  CAMELS was also used in \cite{pandey_2023} to show that tSZ measurements can be used to constrain the impact of feedback on the matter distribution around group-scale halos at $z=0.$ 

As is clear from the varied levels of success in replicating observations, there is still work to be done when it comes to probing AGN feedback with the tSZ effect. Here, we build on this work by using SIMBA simulation to compare how the tSZ signal from stacked galaxy samples changes for models with and without AGN feedback. We compute the average tSZ signal around massive galaxies at redshifts ranging from 0.2 to 1.5 in order to determine at what angular scales and redshift  AGN feedback has the largest impact on the small-scale microwave background. We compare SIMBA's treatment of AGN against current results from ACT at $z \approx$ 0.5 and 1, and, in anticipation of upcoming results from instruments such as the TolTEC camera on the Large Millimeter Telescope Alfonso Serrano (LMT), we use SIMBA to make predictions of the tSZ signal from stacked galaxies at higher-resolutions.  We also explore a novel approach to quantify the non-spherically symmetric features surrounding our galaxies by stacking moments of the tSZ distribution.

The structure of this work is as follows. In \S \ref{sec2_1} we discuss the SIMBA simulations and their treatment of AGN feedback, as well as our methods of selecting simulated galaxy samples. In \S \ref{sec2_2} we present the observational data we compare against at $z \approx 1$ and $z \approx 0.5$, as well as the galaxy samples from SIMBA we use in these comparisons. In \S \ref{sec3}, we discuss the methods by which we generate the tSZ maps and stack our galaxy samples. In \S \ref{sec4} we present our results in the form of radial profiles, moments, and thermal energy measurements, and in \S \ref{sec5} we discuss the implications of these results in terms of AGN feedback models and future observations of the tSZ effect.

\section{Data}\label{sec2}

\subsection{SIMBA}\label{sec2_1}

We used the SIMBA cosmological galaxy formation simulations to produce maps of the tSZ signal. SIMBA makes use of the meshless finite volume solver of GIZMO but separates itself from its parent code in several ways \citep{gizmo, dave}, including a novel sub-grid prescription for modeling AGN feedback.  SIMBA includes two feedback modes corresponding to different accretion rates for supermassive black holes (SMBH). High accretion rates ($f_{\text{Edd}} \equiv \text{M}_{\text{BH}}/\text{M}_{\text{Edd}} > 0.2$) lead to outflows of molecular and ionized gas moving at velocities $\approx$ 1,000 km  $\text{s}^{-1}$. In this radiative mode, the velocity of wind particles is given by:
\begin{equation}
    v_{w, \text{ Radiative}} = 500+ \frac{500}{3}\left( \text{log}_{10} \frac{M_{\text{BH}}}{M_\odot}-6 \right)\text{ km s}^{-1}.
\end{equation}
Black holes in low Eddington accretion mode ($\text{f}_{\text{Edd}} <0.2$) produce jet-type feedback with velocities given by:
\begin{equation}
    v_{w, \text{ Jet}} = v_{w, \text{ Radiative}}+ 7000 \text{ log}_{10} \left(\frac{0.2}{f_{\text{Edd}}} \right) \text{ km s}^{-1}.
\end{equation}
In order to trigger the jet mode, the black hole mass must be greater than $10^{7.5} M_\odot$. 

These two modes follow the generally observed trends of AGN feedback \citep{Best2012,Heckman2014,Perna2017}, although there is still no consensus regarding the underlying mechanism generating these outflows. Aside from these kinetic feedback models, SIMBA also includes X-ray feedback, which is only operational when the jet mode is active,  $M_* > 10^9 M_\odot,$ and $M_{\text{gas}}/M_{\text{baryon}}<0.2$. The X-ray feedback takes the form of volume heating of the gas in the SMBH's accretion kernel with the heating flux dropping as the square of the distance, including Plummer softening. While non-ISM gas is heated according to the flux at its position, for the pressurized ISM gas half of the X-ray energy is applied as a kick to the particles, while the other half is supplied as heat. The X-ray feedback in SIMBA has been shown to have a clear impact on the quenching of galaxies, although it has only a minimal effect on the galaxy mass function \cite{dave}.

For the purposes of this study, we used the publicly available 50 $h^{-1}$ Mpc SIMBA runs. These assume a $\Lambda$CDM model with a Planck Collaboration concordant cosmology of $H = 68$ km/s/Mpc and total matter, vacuum, and baryonic densities respectively  $\Omega_m = 0.3$, $\Omega_\Lambda = 0.7$, and $\Omega_b = 0.048$ in units of the critical density, as described in \cite{2016A&A...594A..13P} and \cite{dave}.   To isolate the effects of AGN and compare them with current data, we focused on two such runs, one with all feedback on (labeled AGN) and a second run where all forms of feedback are turned off (labeled N-AGN). 

\begin{figure}[ht!]
    \centering
    \includegraphics[width = 0.98\linewidth]{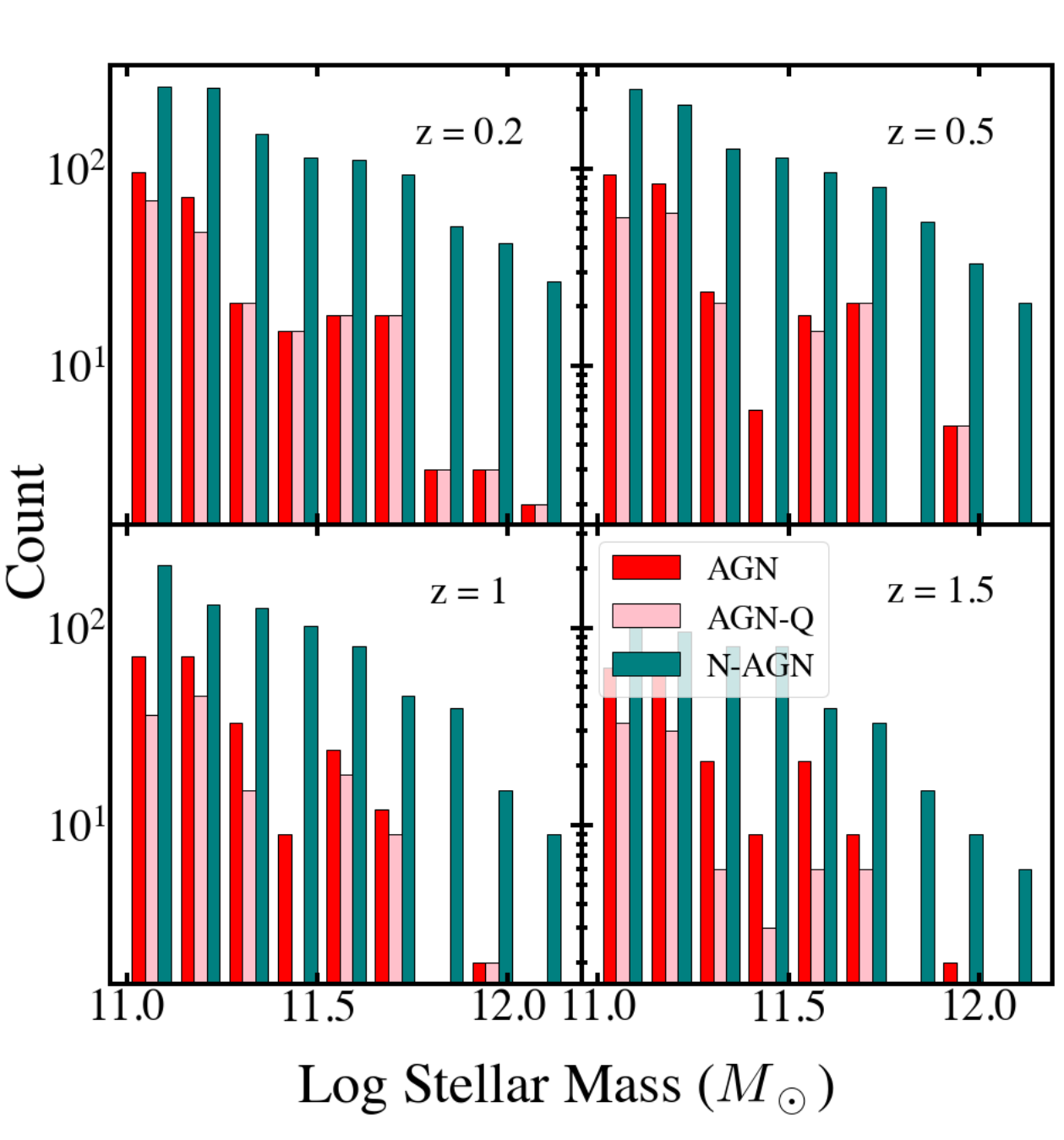}
    \caption{Stellar mass distribution for our AGN (red), quiescent AGN (pink), and No AGN (teal) galaxy samples.  \label{galsamp}}
\end{figure}

In order to identify structures, SIMBA uses a 3D friends-of-friends halo finder written into GIZMO, with a linking length of 0.2 times the mean inter-particle spacing. For both the AGN and N-AGN runs, we used the yt-based package CAESAR to generate a galaxy catalog using 6D friends-of-friends with a linking length of 0.0056 times the mean inter-particle spacing. We also used CAESAR to cross-match galaxies and halos and output a single catalog for analysis  \citep{dave}. 

In order to maximize the sample size, we projected the box in three directions. While the halos are expected to be spherically symmetric within the virial radius, the behavior at large radii will vary for different projection angles. This is especially important when considering larger beams such as those at ACT and SPT where very little data comes from within the virial radius. 

 We focused our analysis on three different galaxy samples. The first sample, labeled AGN, consists of galaxies with $M_{\text{stellar}}>10^{11} M_\odot$ from the AGN run. These are of particular interest because the energy input from AGN into the medium surrounding them is expected to be large as compared to the energy input from gravitational collapse. Additionally, galaxies smaller than $10^{11} M_\odot$ show substantial growth below $z \approx 1$ and thus are less suitable for studying downsizing \citep{Treu_2005}.

The second galaxy sample, labeled AGN-Q,  is a subset of this group, consisting of quiescent galaxies without ongoing star formation.  It is advantageous observationally to only consider quiescent galaxies to avoid contamination of tSZ signals by AGN radio emission.  As AGN are hypothesized to play a large role in galaxy quenching, quiescent galaxies should also have been significantly affected by AGN feedback. To define this sample, we cut on age and sSFR as in \cite{jeremy} and \cite{meinke_2023} so that the AGN-Q sample has galaxies with $M_{\text{stellar}}>10^{11} M_\odot$, mass-weighted mean stellar age $>1$ Gyr, and sSFR $<$ 0.01 Gyr$^{-1}$. 

The third galaxy sample, labeled N-AGN, comes from the run without feedback. Here, we could not make a cut on quiescent galaxies, as without AGN feedback all massive galaxies continue to form stars at late times. Thus, the only cut made on the N-AGN sample was by stellar mass, setting the same lower limit of $M_{\text{stellar}}>10^{11} M_\odot$ as in the other samples. Additionally, we set an upper limit on this sample, as without AGN feedback there is a significant population of massive galaxies ($M_*>10^{12} M_\odot$) that does not agree with the observed galaxy stellar mass distribution. Thus we cut any galaxies from the N-AGN run that had stellar masses above the most massive galaxy in the AGN run. We recognize that the N-AGN run from SIMBA is not designed with direct comparisons with galaxy surveys in mind. However, we chose to make this cut so as to potentially model how galaxies of a certain stellar mass range might look without the feedback of the type implemented by SIMBA, and to make more one-to-one comparisons with the AGN run. The stellar mass distribution for our three galaxy samples at each redshift is shown in Figure \ref{galsamp}.

\subsection{Observational Data} \label{sec2_2}

\begin{figure*}[ht!]

\centering
{\includegraphics[width=0.45\textwidth]{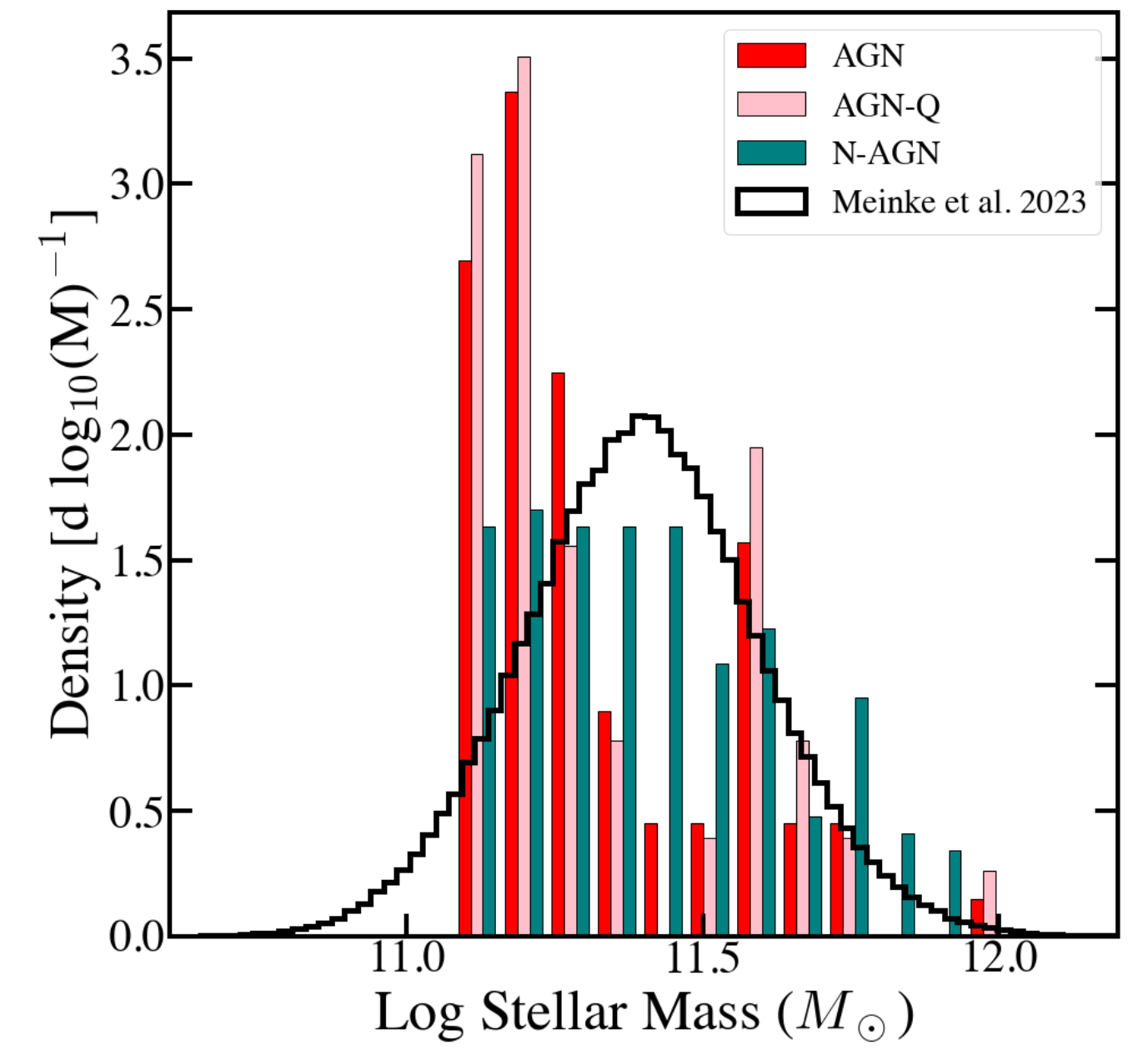}}
\hfill
{\includegraphics[width=0.45\textwidth]{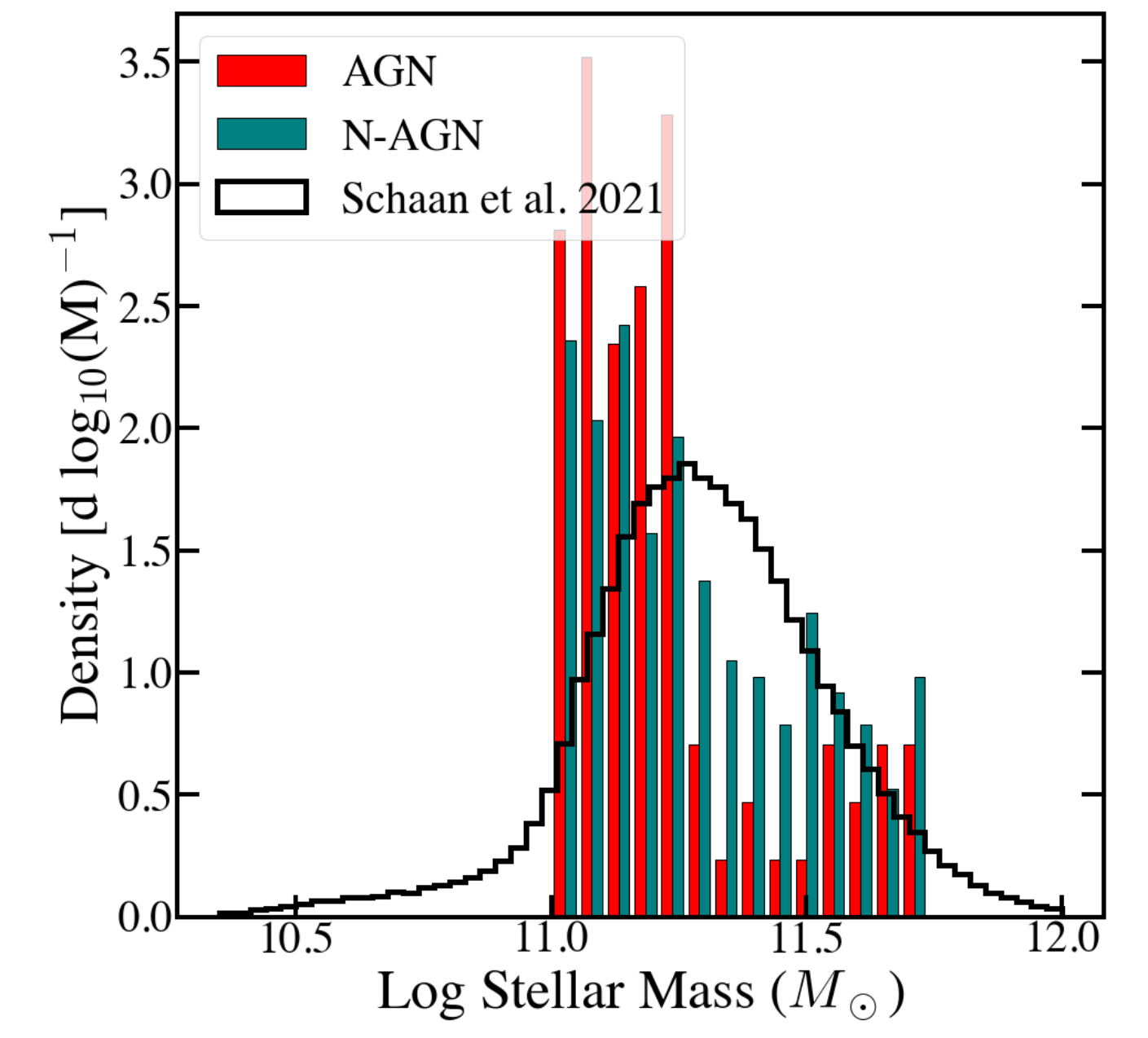}}
\caption{Stellar mass distribution of the SIMBA samples used for observational comparison. Galaxy samples were chosen to match the mean stellar stellar mass. Left: Sample at z=1. SIMBA galaxies are cut to exclude any with  $M_*<1.3 \times 10^{11} M_\odot$. Observational data from \cite{meinke_2023} is shown in black. Right: Sample at z=0.5. SIMBA galaxies are cut to exclude any with $M_*>5.5 \times 10^{11} M_\odot$. The distribution of the observational data from \cite{Schaan_2021} is shown in black. \label{gal_sample_comps}}
\end{figure*} 

While the samples outlined above were used to make predictions for various beam sizes and redshifts, we also compared the results from SIMBA against existing observations of the tSZ effect. At $z \approx 1$, we compared against ACT data as analyzed in \cite{meinke_2023}. The galaxies in this sample were selected from the Dark Energy Survey and Wide-Field Infrared Survey Explorer and have a mean stellar mass of $2.51 \times 10^{11} M_\odot$. To compare against this dataset, we used a sample of SIMBA galaxies taken from three snapshots at redshifts 0.993, 1.114, and 1.211 to match the mean redshift in \cite{meinke_2023} of $z \approx 1.1$ as well as the range of redshifts used. We then extracted three galaxy samples: AGN, AGN-Q, and N-AGN as above, except we made a more precise cut on stellar mass in order to match the mean stellar mass of the observational sample. This meant the exclusion of any galaxies with $M_*<1.3 \times 10^{11} M_\odot$. 

At $z \approx 0.5$, we compared against ACT data from \cite{Schaan_2021}. This work used galaxies selected from the Baryon Oscillation Spectroscopic Survey CMASS catalog (DR10 and DR12), which have an average redshift of z = 0.55 and an average stellar mass of $1.86 \times 10^{11} M_\odot$ \citep{10.1093/mnras/stt1424, Schaan_2021}. To make the comparison with SIMBA data, we used three snapshots at redshifts 0.490, 0.534, and 0.603 to match the mean and approximate range of redshifts of the observational sample \citep{Schaan_2021}. We extracted a sample of galaxies from SIMBA run with all its AGN feedback on (labeled AGN) and SIMBA run with all its AGN off (labeled N-AGN). To match the mean stellar mass of the AGN sample to the mean stellar mass of the observational sample, we cut any galaxies with a stellar mass above $5.5 \times 10^{11} M_\odot$. While generating these samples, we chose to match the mean values over the median as the tSZ signal is determined via stacking, which is sensitive to outliers. As the SIMBA catalog is not a magnitude-limited sample, the distribution of masses will not follow the same approximately normal distribution of the observational sample, so matching the mean value is also preferable to matching the range. This does mean that there is an overabundance of galaxies in the SIMBA population around $1.25 \times 10^{11} M_\odot$ and no SIMBA galaxies in the smallest range of stellar masses at both redshifts. Despite the difference in the distributions, we are assuming that the mean stellar mass has a direct relationship with the signal due to the averaging process inherent in stacking. However, even if this is not the case, the range of masses is small enough that there would be minimal impact on the results. The stellar mass distribution for each of the observational data sets as well as the SIMBA data used to compare against them is shown in Figure \ref{gal_sample_comps}.

\begin{figure*}[ht!]

\centering
{\includegraphics[width=0.47\textwidth]{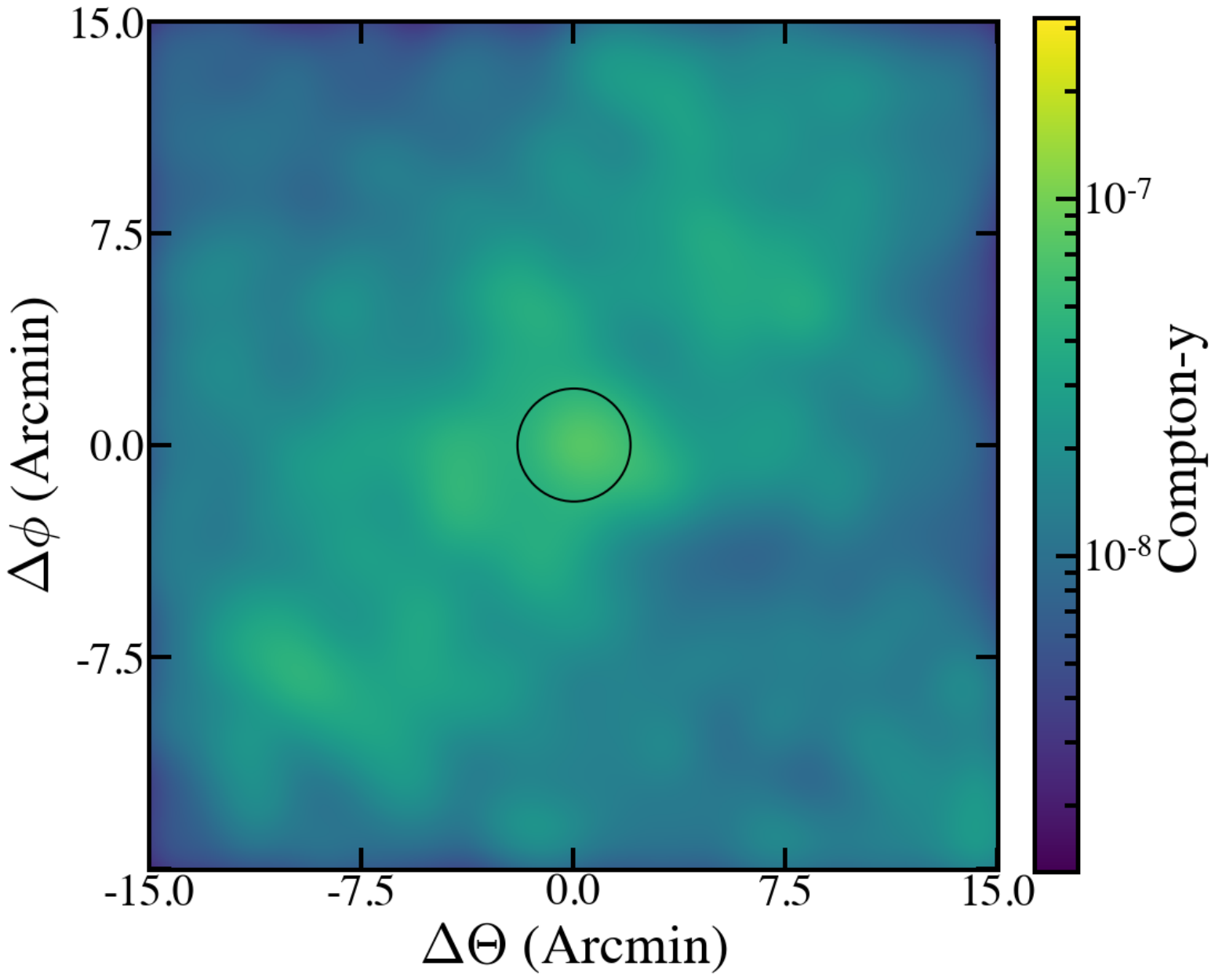}}
\qquad
{\includegraphics[width=0.47\textwidth]{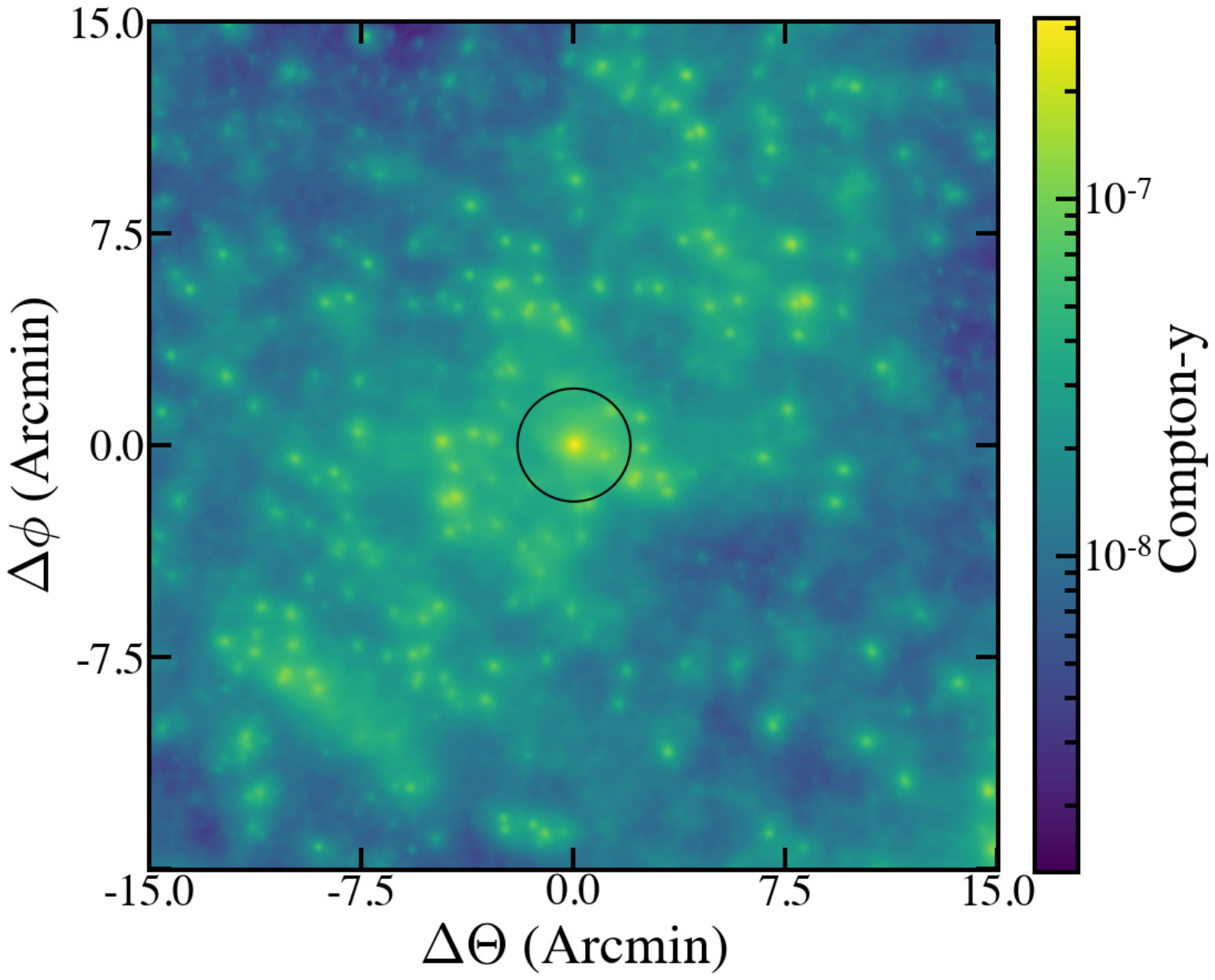}}
\caption{Examples of stacked tSZ y maps for 165 galaxies larger than $10^{11} M_\odot$ from the AGN sample at $z=1.$ The shown regions are 15 arcminutes across, and the black circle represents a region with a 2 arcminute radius used to calculate the thermal energy. We show how the stacks differ when convolved with two different beams. Left: ACT beam, FWHM = 2.1' at 90 GHz. Right: TolTEC beam, FWHM = 9.5'' at 150 GHz. \label{beam}}
\end{figure*} 

\section{Analysis} \label{sec3}

\subsection{tSZ maps}

The initial analysis of the simulation snapshots was conducted using the Python package yt \citep{yt} which represents physical quantities as fields varying across multidimensional space. This includes fields that are already output by the code (i.e. gas temperature and density) as well as derived fields, including the tSZ signal. However, we elected not to use the derived field and calculate the tSZ signal ourselves, to allow us to exclude specific particles.  Allowed us to cut interstellar medium (ISM) particles which are artificially pressurized to $n_H > 0.1 \text{cm}^{-3}$. This pressurization is implemented to prevent runaway star formation in the underresolved ISM of  simulated galaxies as described in \cite{10.1046/j.1365-8711.2003.06206.x}, but it leads to unphysically high tSZ signals.  Additionally, we removed current wind particles that are hydrodynamically decoupled and thus do not have physically-meaningful properties. These cuts remove about 3.5\% of the gas particles from each snapshot.

After removing these particles, we generated maps of the tSZ-$y$ signal from the gas density and temperature fields according to Eq.\ (\ref{eq1}). The maps had a pixel size of 2.5 arcseconds, which was set to be at a higher-resolution than any current millimeter-wave optics. Setting this high resolution allowed us to convolve the maps with a range of beams for observational comparisons. The pixelization process for the line of sight integration needed to calculate the tSZ-$y$ signal is done in yt via the software SPLASH \citep{2007PASA...24..159P}. SPLASH calculates the projected value of an arbitrary quantity $A$ at point ($x$,$y$) with 
\begin{equation} 
A(x,y) = 
 \sum\limits_{j} \frac{m_j}{\rho_j} A_j Y(x-x_j,y-y_j,h_j),
\end{equation}
where  $m_j$ and $\rho_j$ are the mass and density of particle j, $A_j$ is the quantity of A defined on particle j, and $Y$ is the 3D cubic spline kernel integrated through one spatial dimension \citep{2007PASA...24..159P}. While this method will consider the value of the kernel in the center of each pixel, that is not necessarily the average over the pixel, leading to the possibility of slight errors in the projection. This could lead to issues such as lack of mass conservation if pixels are large, and if the center of the pixels does not line up with the particle positions, signal can be lost. In our case, pixel sizes are about 1/5 the typical smoothing length of the gas particles, so errors due to the pixelization process are minimal. As we later convolve with a larger beam for our observational comparisons, the difference in resolution between the smoothing length and pixelated map is negligible.

\subsection{Beam Effects}

The entire map of the SZ-$y$ signal was convolved with a gaussian beam, using the Fast Fourier Transform mode in the astropy package \citep{2013A&A...558A..33A, 2018AJ....156..123A, 2022ApJ...935..167A}. The first beam we used corresponded to the lowest resolution data from the Atacama Cosmology Telescope (ACT), with a FWHM resolution of 2.1' in the band centered at 90 GHz. This is the limiting resolution for the prior SZ analysis from ACT data as seen in \cite{meinke_2023}. The second beam we used was for the newly installed TolTEC camera, which was mounted on the LMT in the Spring of 2022 and began commissioning in the following summer. The lowest resolution for TolTEC is 9.5'' at 150 GHz \citep{toltec_prec}. While not at precisely the same frequency bins, these values represent the lowest resolution from each instrument, allowing us to make conservative estimates on spatial resolution going forward. 

A visual comparison of the two beams for the AGN simulation projected in one direction is shown in Figure \ref{beam}. It is apparent that neighboring galaxies will have a significant impact on the signal, particularly with the higher-resolution TolTEC beam. We generated and convolved maps for SIMBA snapshots at four different redshifts: 0.211, 0.490, 0.993, and 1.497,  which we simplify to 0.2, 0.5, 1, and 1.5 for the rest of the paper. 

We stacked the Compton-$y$ signal at each redshift for the galaxy samples defined in \S \ref{sec2} by averaging around the center of mass determined using the CAESAR catalogs.

\section{Results}\label{sec4}

\subsection{Radial Profiles}

One approach for distinguishing feedback models is analyzing the radial profile of the Compton $y$ signal. To subtract any large-scale Compton signal unassociated with the galaxies of interest, we calculate the average signal in the radial bins from 15-20 arcminutes for each beam and subtract it from the entire radial profile. This method of background subtraction was used in the observational work we compare to a redshift of one \citep[see][]{meinke_2023}. It also is worth noting that as the box is only 50 $h^{-1}$ comoving Mpc across, the behavior at large radii (above $\approx 10'$) will be affected by the suppression of structures near a significant fraction of the box-scale. However this will not have an impact on CGM or feedback behavior at small radii, where we focus our analysis.

We used a bootstrapping approach to quantify both the error in each a radial bin and the correlations between bins. To calculate the error in each bin, we used SciPy to generate 4,000 bootstrapped samples and determined the error in the distribution of those samples' means \citep{2020NatMe..17..261V}.  To understand the correlation between bins, we generated 4,000 bootstrapped samples of our galaxies and created radial correlation matrices for both the ACT and TolTEC beam, as shown in Appendix \ref{appa}. 

\begin{figure}[t]
    \centering
    \includegraphics[width = 0.49 \textwidth]{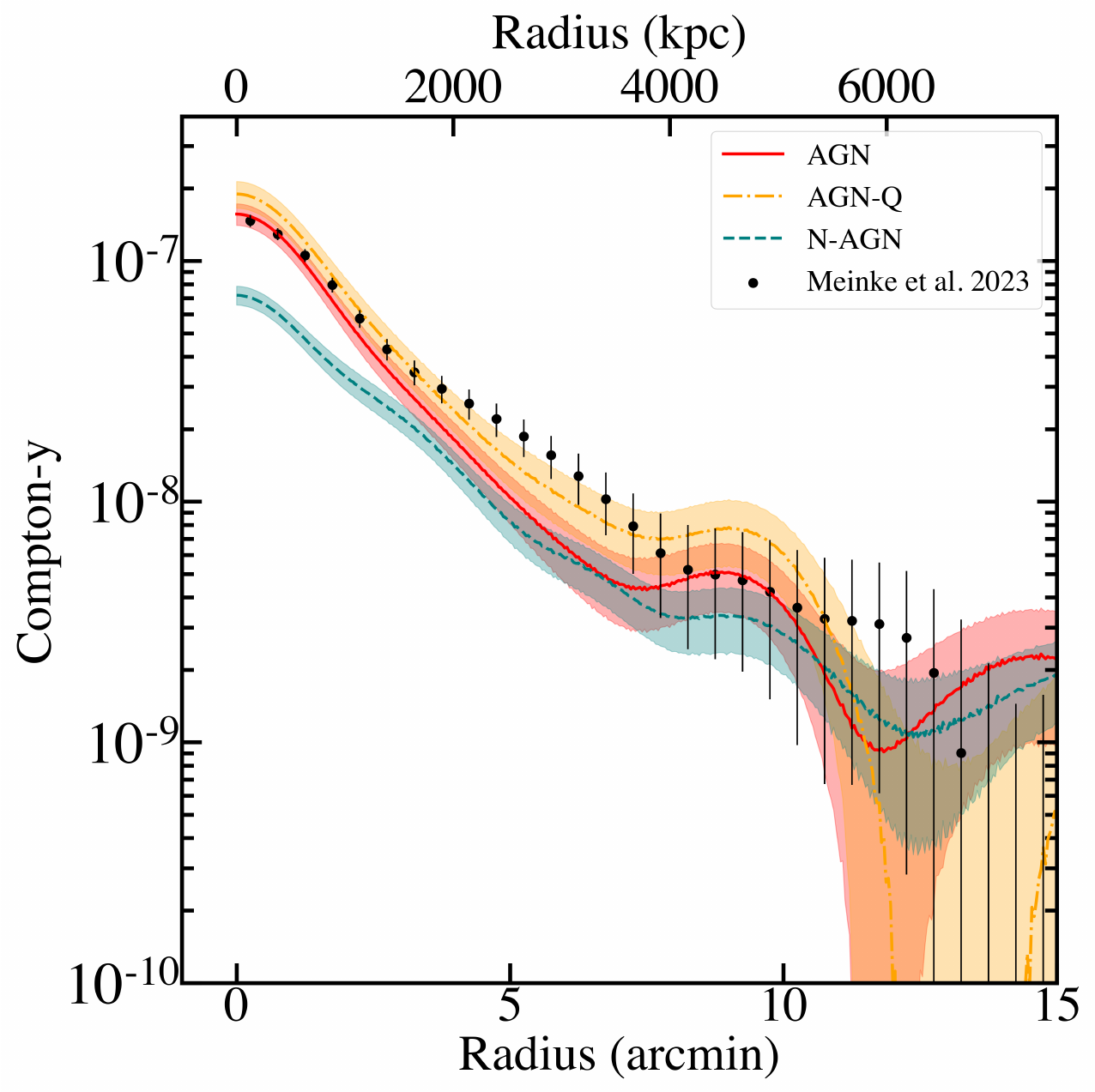}
    \caption{Compton-y radial profiles at $z \approx 1$ for stacked galaxies convolved with a 2.1' beam to compare against observational data from \cite{meinke_2023} (black). The kpc scale on the upper x-axis is calculated using the mean z of the three snapshots ($\approx 1.1$). We see strong agreement between the observations and the AGN-Q galaxies from SIMBA. Shaded regions represent 1-$\sigma$ errors on the SIMBA results.}
    \label{actcomp}
\end{figure}

\subsubsection{Observational Comparisons}

Figure \ref{actcomp} shows the results for z $\approx$ 1 data from three snapshots convolved with the 2.1' beam compared against ACT data from \cite{meinke_2023}. In order to match the observational sample as closely as possible, we cut any galaxies with $M_*<1.3 \times 10^{11} M_\odot$ so as to align with the average stellar mass of the observational sample ($2.5 \times 10^{11} M_\odot$) as described in \S \ref{sec2_2}. To quantify the agreement between SIMBA and ACT at $z \approx 1$, we calculate the number of pooled standard deviations ($\sigma = \sqrt{\sigma_1^2+\sigma_2^2}$) between the $y-$signals in each radial bin. This is done by taking the difference between the results and dividing by the error of each result added in quadrature. These results are summarized in Table \ref{sigma}, where we present the separation between observational and simulated results at various radii. 

We see that the AGN-Q sample is within 1.5$\sigma$ from observational data across the whole profile, although SIMBA predicts a slightly higher signal than what is observed at the very center of the stack. 
We can also see that the observational data is clearly inconsistent with the N-AGN run, providing evidence that the galaxies observed with ACT have experienced AGN feedback. The generally strong agreement seen at z$\approx$1 gives us the confidence that SIMBA's treatment of AGN lines up well with observation, and we can use SIMBA to make predictions about higher-resolution data at this redshift. 

\begin{figure}[t]
    \centering
  \includegraphics[width = 0.49 \textwidth]{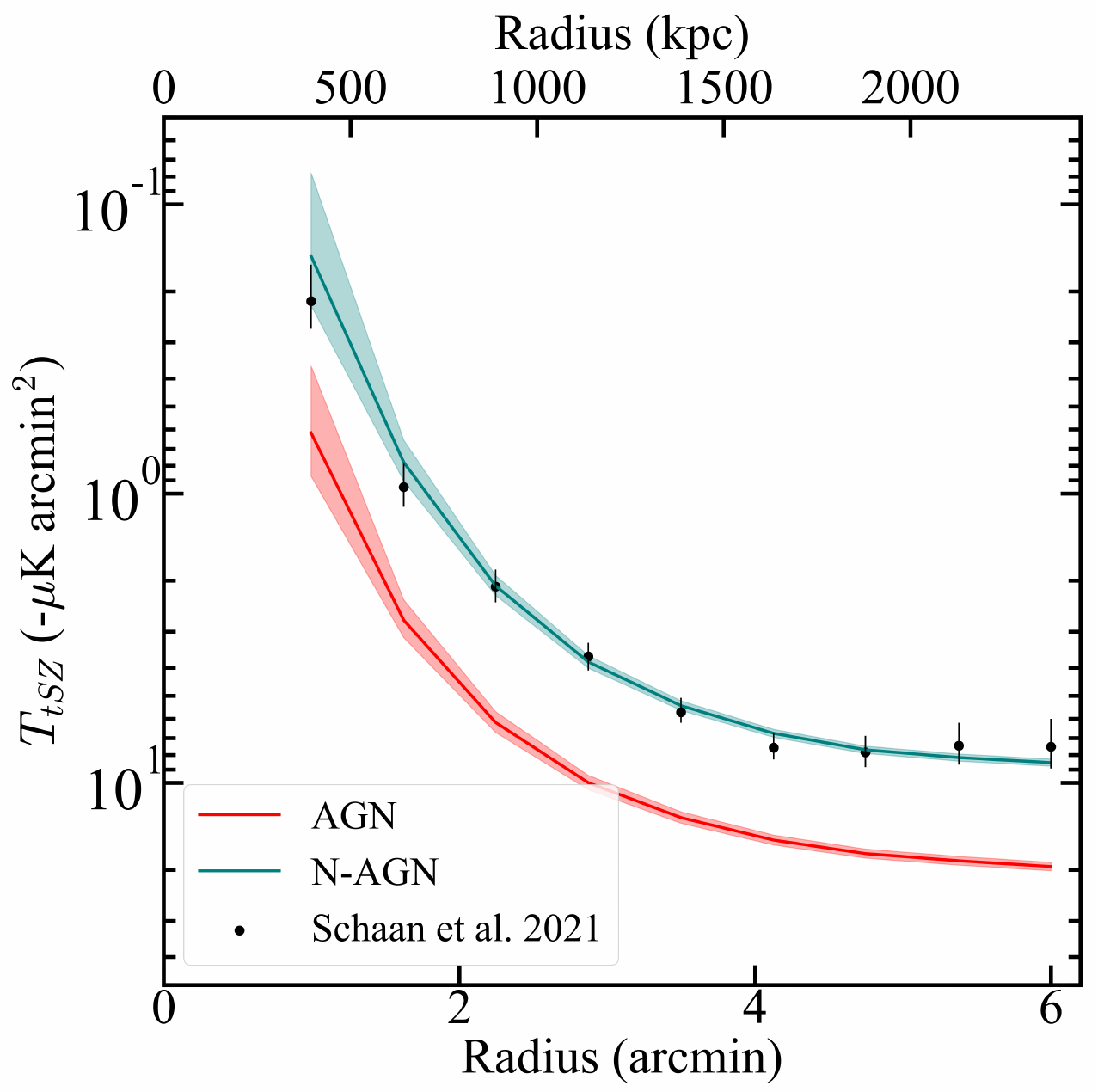}
    \caption{Radial profiles of the temperature change in the CMB at 150 GHz due to the tSZ effect at $z \approx 0.5$ for stacked galaxies convolved with a 2.1' beam and CAP filter to compare against observational data from \cite{Schaan_2021} (black).  The kpc scale on the upper x-axis is calculated using the mean z of the three snapshots ($\approx 0.55$). At this lower redshift, we see less of an agreement between SIMBA and observation. Shaded regions represent 1-$\sigma$ errors on the SIMBA results.}
    \label{schaancomp}
\end{figure}
\begin{figure*}[ht!]

\centering
\includegraphics[width=0.9\textwidth]{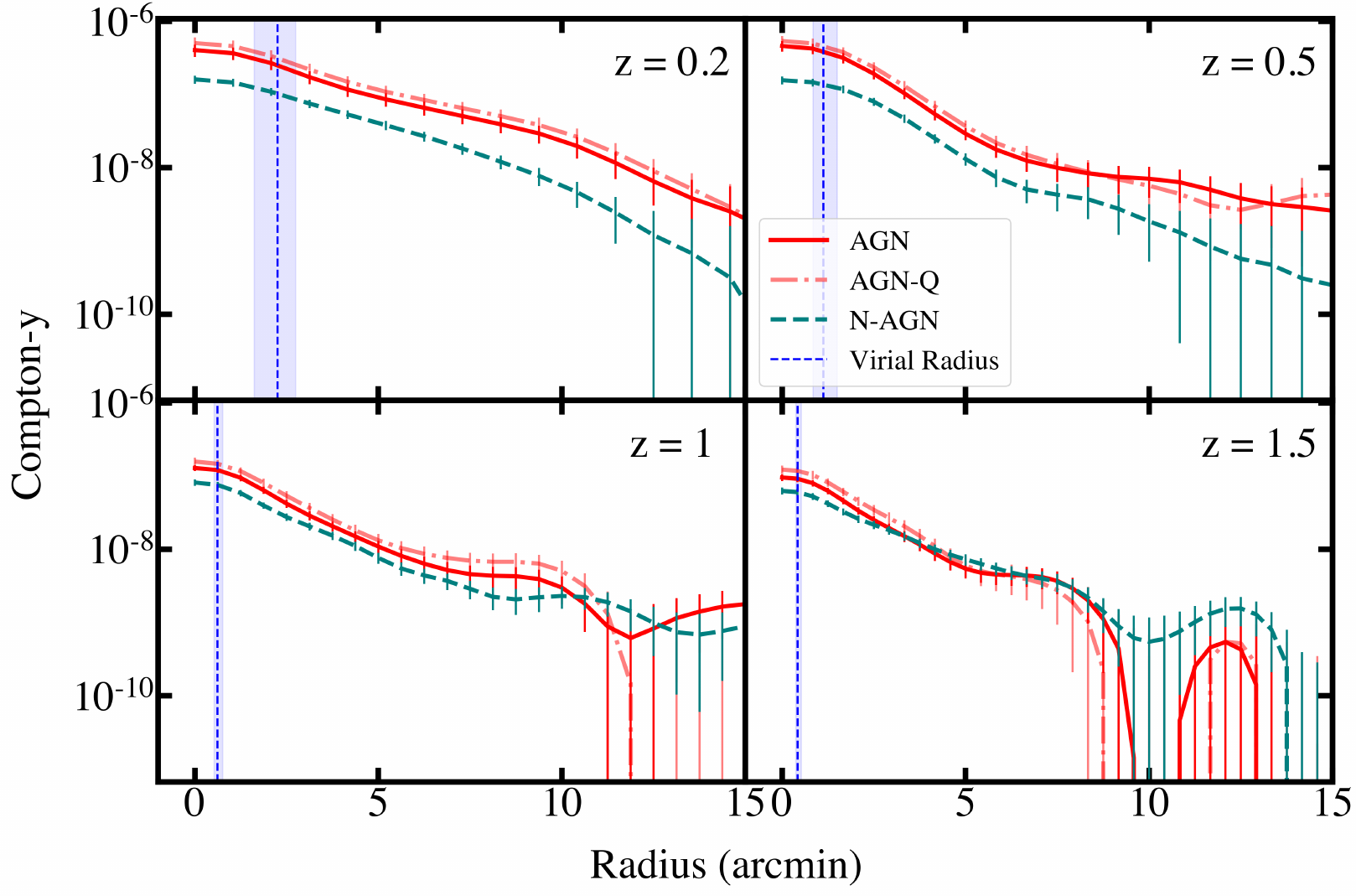}
\vspace{-0.1in}
\caption{Compton-y radial profile as a function of redshift for SIMBA data convolved with a 2.1' beam. Shown are the radial profiles for massive galaxies run with AGN feedback, quiescent and massive galaxies with AGN feedback (AGN-Q), and massive galaxies run without AGN (N-AGN). The vertical blue line shows the median virial radius for the AGN galaxy sample, with shaded regions representing the first and third quartile. The median radius in kpc is  480, 420, 300, and 220 for z=0.2, 0.5, 1, and 1.5 respectively. \label{radialact}}
\vspace{0.15in}
\end{figure*} 

Figure \ref{schaancomp} shows the results of three SIMBA snapshots at $z \approx 0.5$ compared with ACT data from \cite{Schaan_2021}. The SIMBA results consist of galaxies with stellar masses $<5.5 \times 10^{11}M_\odot$. This cut was chosen to match the mean stellar mass of the observational sample, as described in \S \ref{sec2_2}. 

The results are shown in Figure \ref{schaancomp}, in which the data takes the form of $T_{tSZ} = \Delta T$ from Eq.\ (\ref{overview_tsz}) at 150 GHz. The radial profiles are plotted using a CAP filter, where the signal at a given radius is found by integrating the signal interior to the radius and subtracting out an integrated ring encompassing the same area exterior to the radius. bAs opposed to the results at $z \approx 1$, we do not see a good agreement between the $z \approx 0.5$ observations and the SIMBA run with AGN feedback. The N-AGN run is much more consistent, with an average separation of  $0.54 \sigma$ as compared to $7 \sigma$ for the AGN sample. While it is possible that this suggests the \cite{Schaan_2021} data does not have evidence of AGN feedback, it is more likely that SIMBA's treatment injects too much energy, particularly at lower redshifts, between $z$ = 1 and 0.5.

\begin{figure*}[ht!]

\centering

\includegraphics[width=0.9\textwidth]{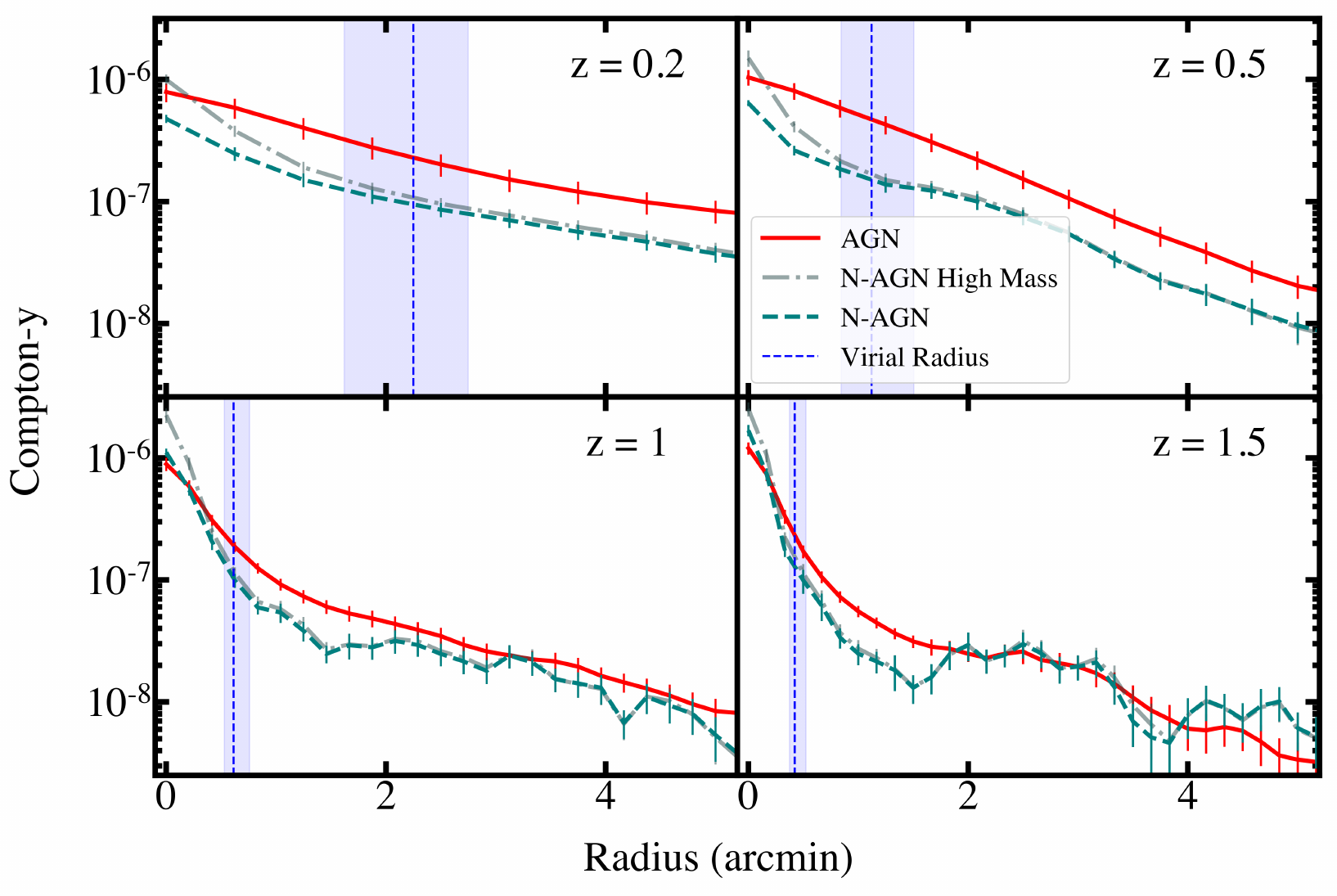}
\vspace{-0.1in}
\caption{Compton-$y$ radial profile at small radii for SIMBA data convolved with a 9.5'' beam. Shown are the radial profiles for massive galaxies run with AGN feedback (red), and massive galaxies run without AGN (teal). Also shown in grey is the N-AGN sample without making the upper cut on stellar mass. While this is an unphysical galaxy population, we show it to demonstrate why the behavior at high redshifts differs from that at low. The vertical blue line shows the median virial radius for the AGN galaxy sample, with shaded regions representing the first and third quartile. \label{radialt2}}
\vspace{0.15in}
\end{figure*} 

\begin{table}[]
    \centering
    \caption{Observational Comparison at z=1}
    \begin{tabular}{c c c c }
    \hline
    \hline
         Radius & AGN-Q & AGN  & N-AGN\\
         
         \hline
         Center & 1.534 $\sigma$ & 0.332$\sigma$ &  7.19$\sigma$\\
         2' & 0.429$\sigma$ &  1.144$\sigma$ & 4.498$\sigma$\\ 
         5' & 1.137$\sigma$& 2.346$\sigma$ & 3.112$\sigma$\\ 
         10' & 0.007$\sigma$ & 0.469 $\sigma$ & 0.439$\sigma$\\ 
         15' & 0.386$\sigma$ & 1.139$\sigma$ & 1.054$\sigma$\\ 
         \hline
    \end{tabular}
    \tablecomments{The differences in $y$-signal are expressed in terms of the pooled standard deviation, $\sigma$. }
    \label{sigma}
\end{table}

\subsubsection{Predictions by Beam and Redshift} \label{radial_predictions}

Figure \ref{radialact} shows the redshift dependence of the radial profiles for SIMBA data of our full galaxy samples as defined in \S \ref{sec2_1} convolved with a 2.1' beam. In order to quantify the differences between runs with and without AGN feedback we once again compute the difference of the signals divided by the pooled standard deviation, as summarized in Table \ref{sigmat}.  Although feedback should have the largest impact on the CGM within two times the virial radius, smaller than observations with this beam can resolve, we do see differences greater than 2$\sigma$ for $z$ $\leq 1$, with the largest difference found in the central signal at a redshift of 0.5. 

\begin{table}[]
    \centering
    \caption{Difference Between AGN and N-AGN}
    \begin{tabular}{ c c c c c c c }
    \hline
    \hline
    \multirow{2}{*}{z} &
      \multicolumn{3}{c}{TolTEC Beam} &
      \multicolumn{3}{c}{ACT Beam} \\
      \cline{2-7}
          & 0'  & 2.5' & 5' & 0'  & 2.5' & 5' \\
         \hline
         
         0.2 & 2.18$\sigma$  & 2.67$\sigma$  & 2.49$\sigma$  & 2.18$\sigma$  & 2.95$\sigma$  & 2.47$\sigma$ \\
         
         0.5 & 2.62$\sigma$ & 2.71$\sigma$ & 2.04$\sigma$  & 3.93$\sigma$& 3.41 $\sigma$ & 2.63$\sigma$  \\
         
         1 & 1.59$\sigma$ & 1.27$\sigma$& 1.00$\sigma$& 3.21$\sigma$& 2.34$\sigma$ & 2.31$\sigma$ \\
         
         1.5 & 2.13$\sigma$ & 0.44$\sigma$& 1.11$\sigma$ &  2.74$\sigma$&0.84$\sigma$ & 0.96$\sigma$\\
         
         \hline
    \end{tabular}
    \tablecomments{The TolTEC beam is 9.5'', the ACT beam is 2.1'. Differences are expressed relative to the pooled standard deviation, $\sigma$.}
    \label{sigmat}
\end{table}

Figure 
\ref{radialt2} show the redshift dependence of the radial profiles convolved with TolTEC's smaller 9.5'' beam. While the lower-resolution ACT beam showed AGN feedback having a higher signal  at every redshift, with the higher-resolution data, it becomes apparent that the N-AGN run actually contains a steeper central profile and a larger central high-redshift signal. 

This difference is most likely due to feedback from jets pushing gas to larger radii in the AGN runs, which would otherwise reach higher densities within the central halo. The grey line in Fig.\ \ref{radialt2} shows the radial profile when the cut we make on the maximum stellar mass of the N-AGN sample is removed. This shows that if we allowed the same galaxy population to evolve without making a mass cut for N-AGN, we would see the higher central signal at lower redshifts as well.    

Overall, when working with higher-resolution data, these results predict that observations at $z \approx 0.5-1$ would be the most useful to determine the presence of AGN. At $z=1.5$ it is hard to distinguish between the feedback models, suggesting that higher-$z$ galaxy samples will not be as useful for this work at TolTEC's resolution. It is worth noting that although the ACT beam reveals a larger difference between AGN and N-AGN at $z \approx 0.5$, its overall signal is 2-4 times lower than TolTEC's, making it more difficult to get a significant detection observationally. 

\subsubsection{Moments}\label{moments}

\begin{figure*}[ht!]
\centering
{\includegraphics[width=0.47\textwidth]{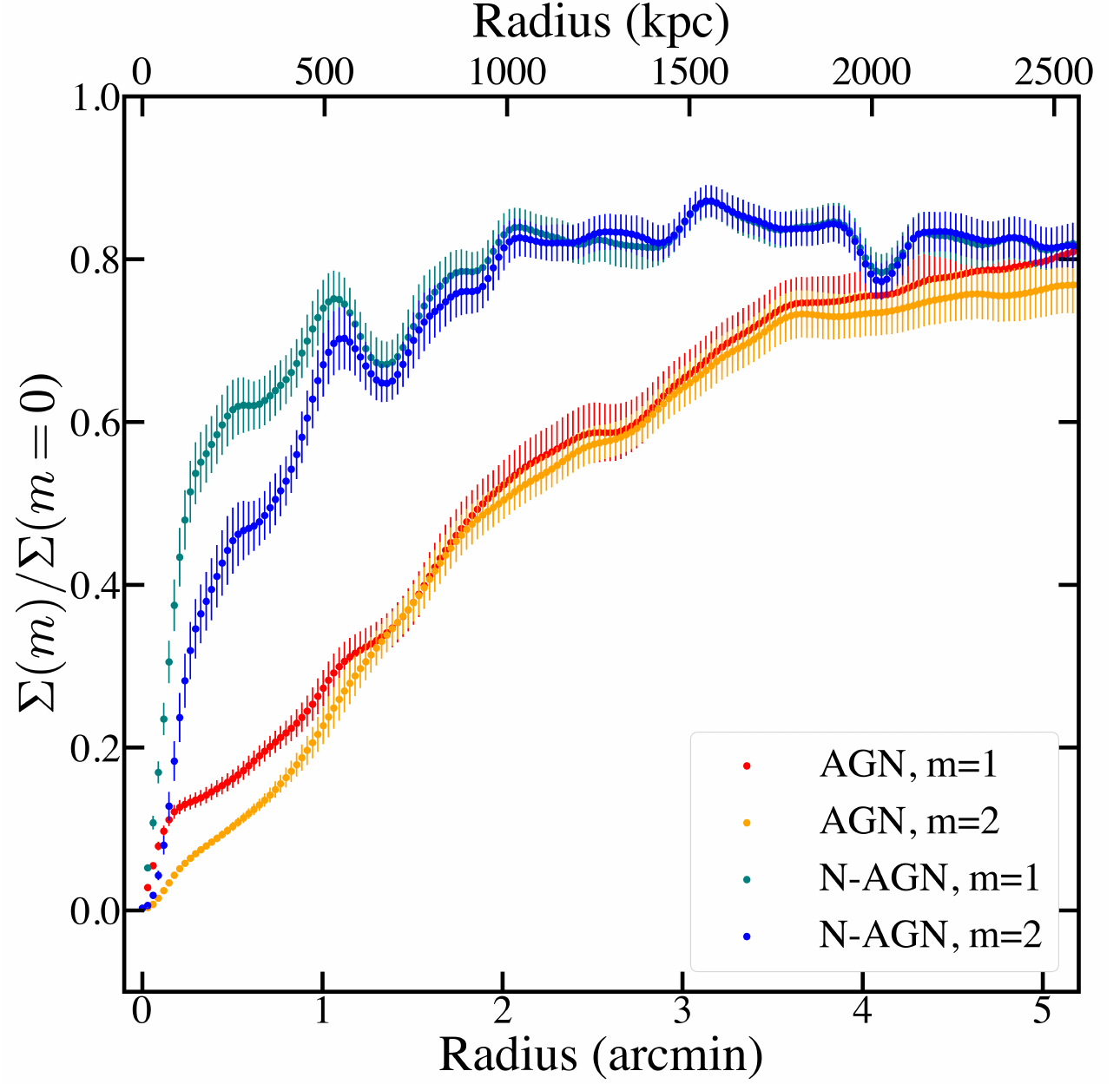}}
\qquad
{\includegraphics[width=0.47\textwidth]{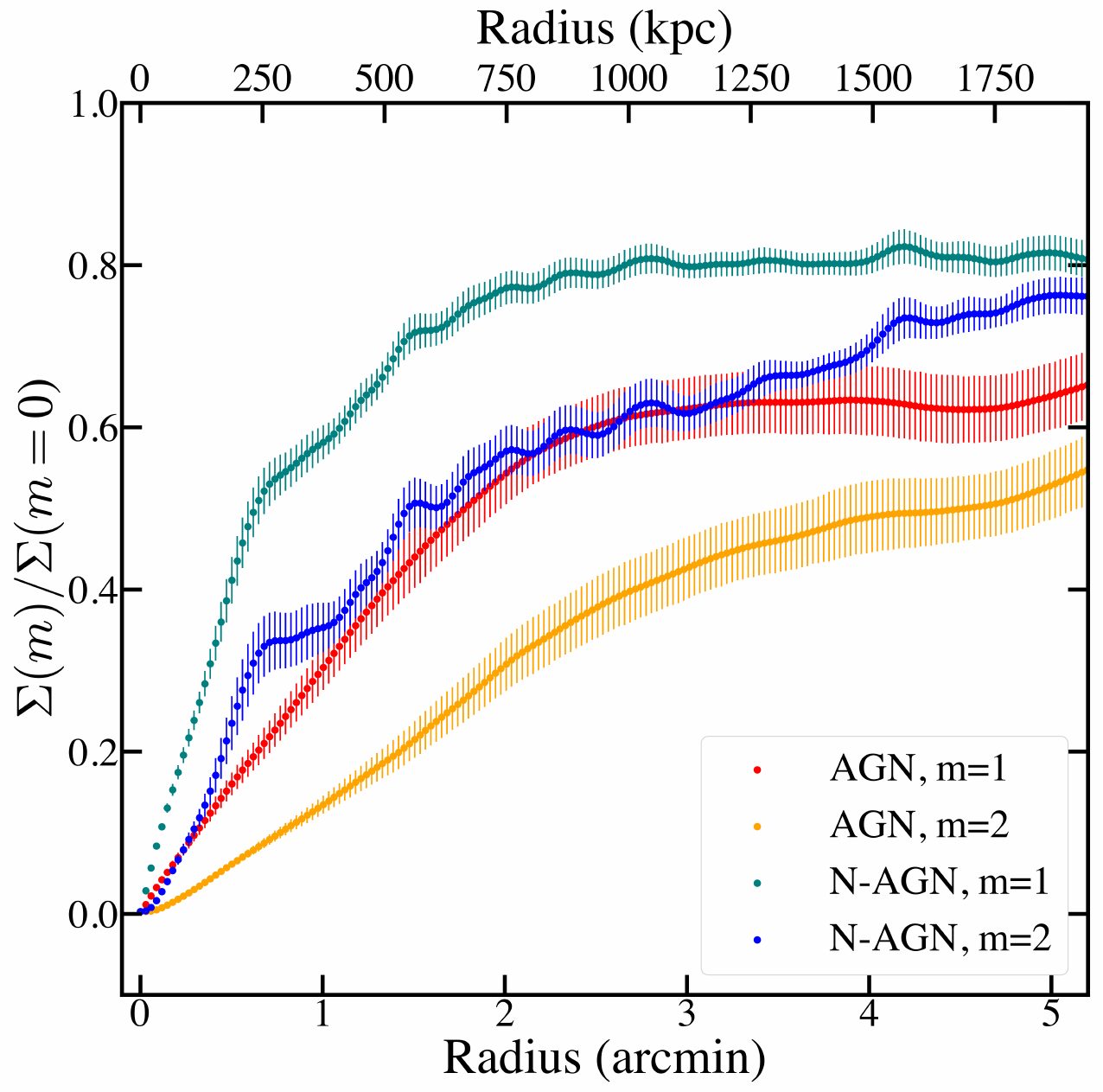}}
\caption{Radial profiles showing the ratio of $\Sigma(m=1)/\Sigma(m=0)$ and  $\Sigma(m=2)/\Sigma(m=0)$ for two galaxy samples at $z \approx 1$ (left) and $z \approx 0.5$ (right). Here, m refers to the number of folds in the symmetry, and higher values represent more asymmetries in the stacked tSZ signal. The lower signals of the AGN stacks can be attributed to the `smoothing' of gas by AGN feedback in SIMBA. The difference between the m=1 and m=2 curves at $z \approx 0.5$ suggests a dipole distribution in the stack, most likely due to galaxies clustering together over time.  \label{momentsplot}}
\end{figure*} 

When generating the stacks to make the radial profiles in \S \ref{radial_predictions}, we are only gathering spherically symmetric data insofar as we take the average signal in each radial bin. Here, we consider a means of extracting the non-spherically-symmetric part of our data. Prior to stacking each galaxy, we transformed $y(r,\theta)$ into 
\begin{equation}
\Sigma(r,m) \equiv \frac{1}{2\pi}\int_0^{2\pi} y(r,\theta) e^{i m \theta} d\theta,
\end{equation}
where $m$ gives information about the number of folds in the symmetry, such that stacking $\Sigma(r,m=0)$ gives the radial profiles above. We generated radial profiles around galaxies from the AGN and N-AGN samples for $m=0$ to 2, quantifying the dipole and quadrupole distribution around the galaxies.

The left panel of Figure \ref{momentsplot} shows the ratio of the higher-moment profiles to the $m=0$ profile for galaxies in the AGN and N-AGN run at $z=1.$ We see similar behaviors for the two higher moments, which is consistent with a singular asymmetry around our galaxies, such as a filamentary structure. The slightly higher signal for $m=1$ at intermediate radii is also consistent with this interpretation. The higher signals for SIMBA ran without AGN can be attributed to SIMBA's treatment of AGN. As seen in the radial profiles above, AGN feedback acts to smooth out the gas distribution, so without it there will be more asymmetries.

\begin{figure*}[ht!]

\centering
{\includegraphics[width=0.45\textwidth]{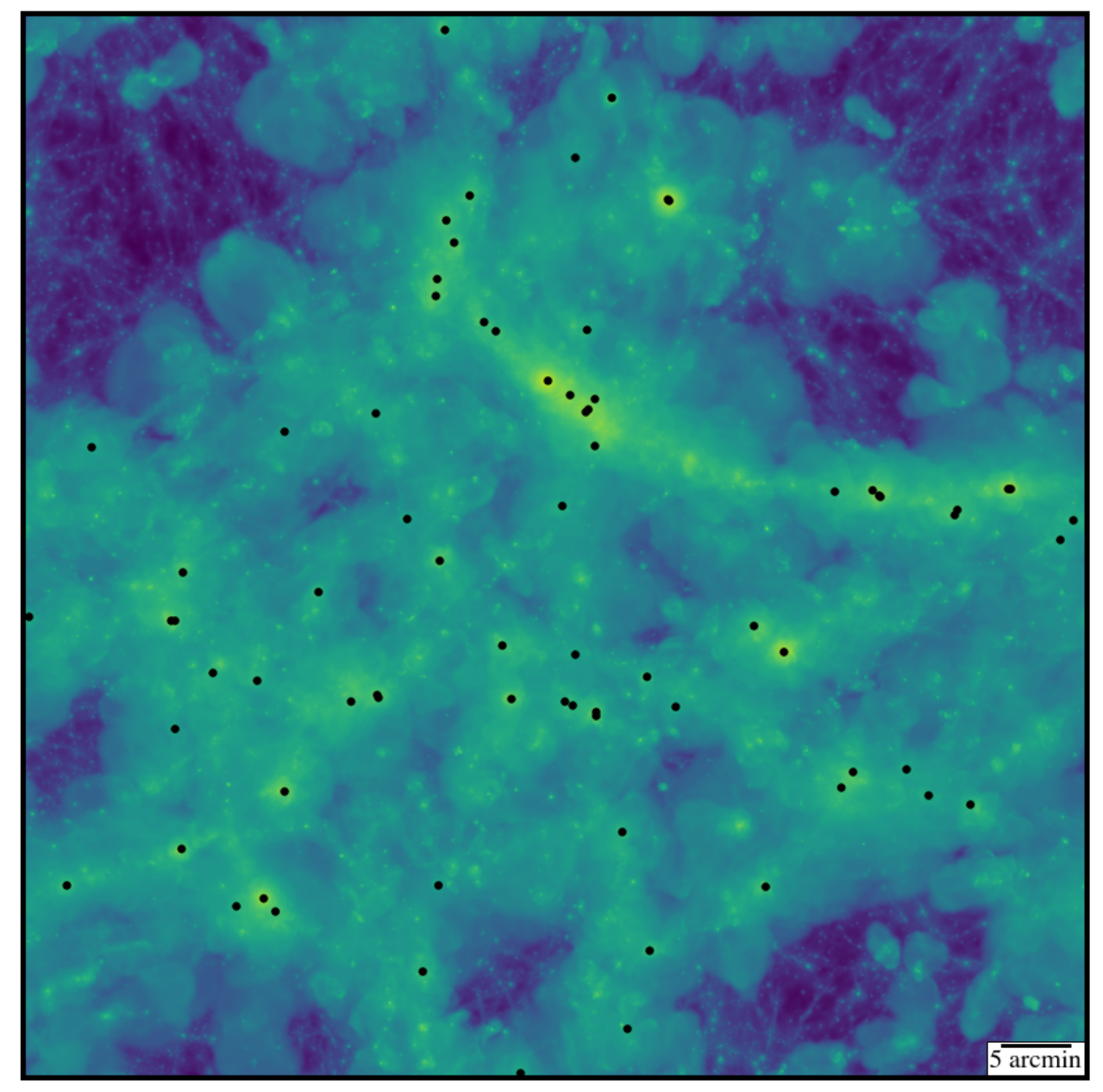}}
\qquad \qquad
{\includegraphics[width=0.45\textwidth]{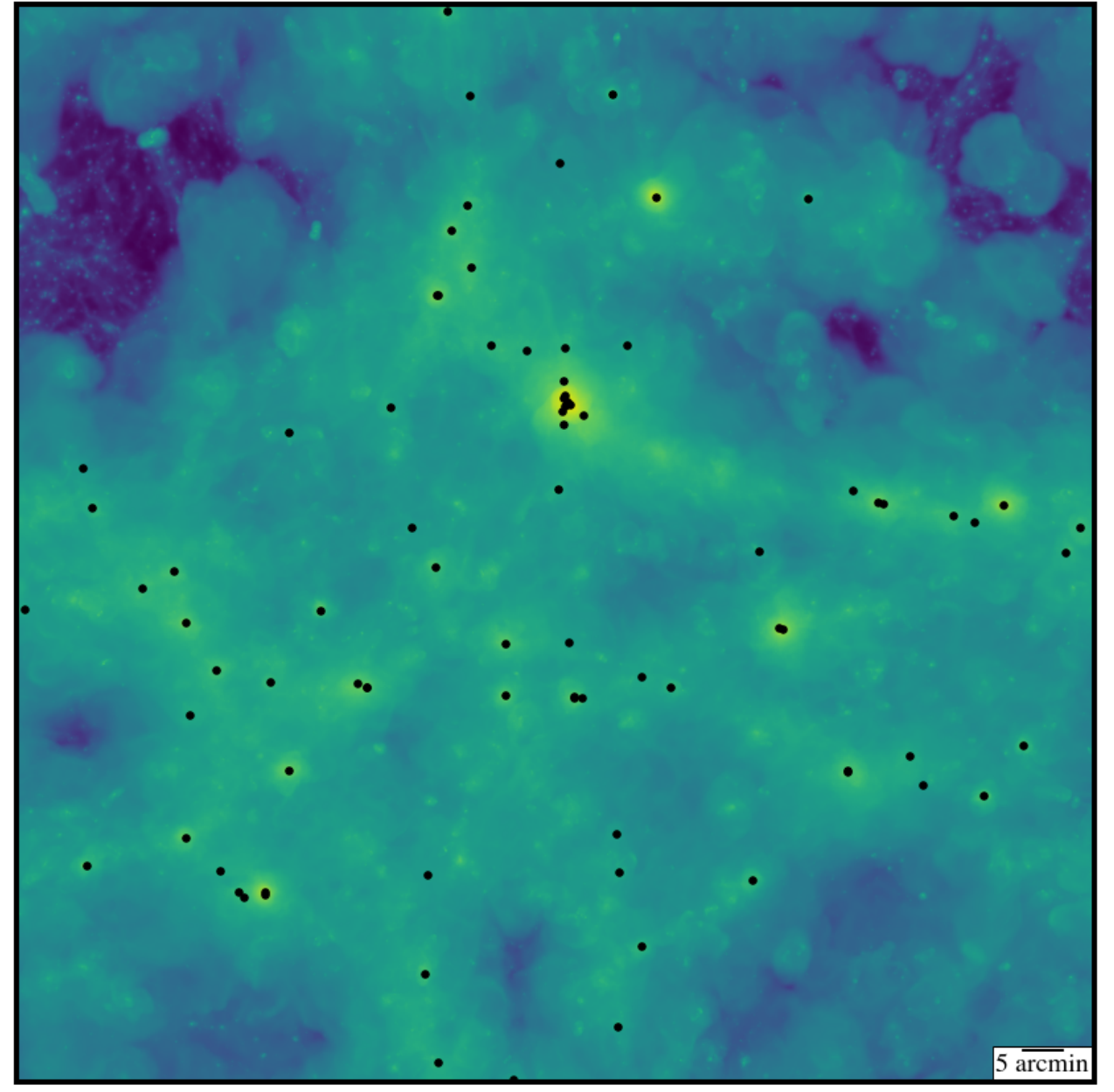}}
\caption{Projected map of the Compton-y parameter at redshift 1 (left) and 0.5 (right) for SIMBA run with all AGN feedback on. The black dots represent the locations of the galaxies used for our analysis. At a redshift of 0.5, the galaxies are more clustered together, largely accounting for the difference in the moments profiles seen in Figure \ref{momentsplot}. \label{galaxy_locations}}
\end{figure*} 

The right panel of Figure \ref{momentsplot} shows the same results but evaluated at a redshift of 0.5. Here we see less of an agreement between the $m=1$ and $m=2$ profiles, with the $m=1$ profile having a significantly stronger signal than $m=2$ at all radii. This is most likely due to galaxies clustering together over time. This leads to more galaxies having a large tSZ signal in their surroundings, which in turn produces a dipole signal in the stack accounting for the m=1 dominance. We also continue to see generally higher signals in the N-AGN sample, showing that the effective ``smoothing" of the gas distribution by AGN continues to late times. Figure \ref{galaxy_locations} shows the distribution of our galaxy sample over a map of the projected tSZ signal. At a redshift of 0.5, we find the massive galaxies are more clustered together, with a significant number of large galaxies being found near the edges of massive halos.

\subsection{Thermal Energy}

Finally, we consider the thermal energy in a small radius around our stacked galaxies. To do this, we fix a circular aperture with a radius corresponding to 2 comoving Mpc as shown by the circles overlaid in Figure \ref{beam}. We picked this size to align with \cite{jeremy} and \cite{meinke_2023}, who argue that this is large enough to contain a majority of the central signal while minimizing noise from neighboring galaxies. In order to capture the same information at each redshift, we changed the radius of the aperture so that it enclosed the same comoving size. At a redshift of $1.0$ the aperture has a radius of 2', while at $z=0.5$ and 0.2, the aperture had a radius of 3.6' and 8.35', respectively. Finally, at a redshift of 1.5,  2 comoving Mpc corresponded to 1.55', which is below the resolution of ACT, so we used the 2' radius at this redshift. 

We computed the signal, expressed as a thermal energy according to equation \ref{eq:therme},
 for galaxies stacked in bins set by both stellar and halo mass at each redshift. When selecting galaxies, we retain the same limits on stellar mass set for each sample in \S \ref{sec3} although we still cut the most massive galaxies from the N-AGN sample as explained above. Uncertainties were determined using a bootstrapping method by resampling  the galaxies in each mass bin with replacement and determining the mean signal for each selection. This process was conducted 4,000 times using the SciPy bootstrap function \citep{2020NatMe..17..261V}. 

\begin{figure}[t!]
\includegraphics[width =  \linewidth]{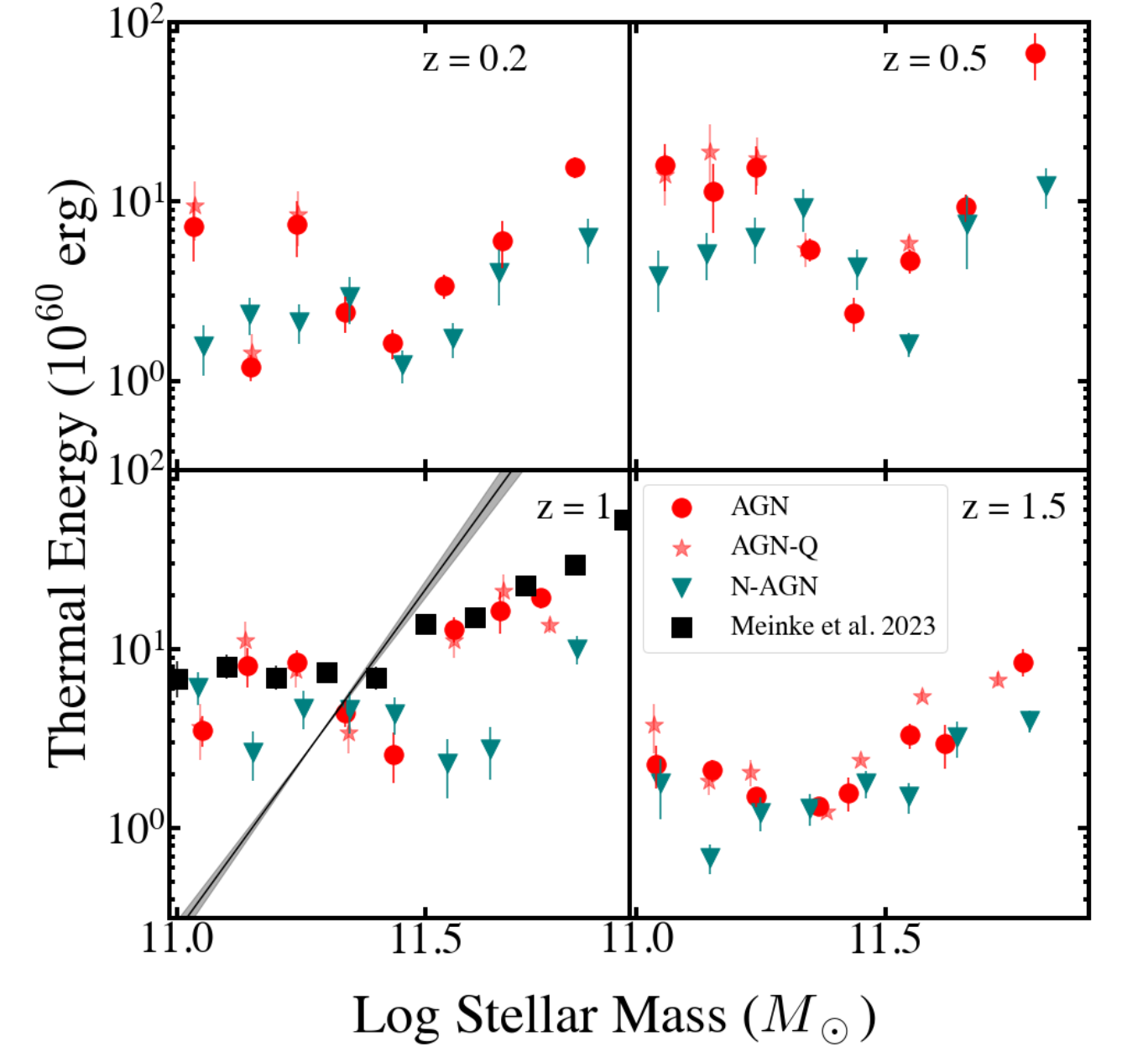}
\caption{Thermal energy measured within a 2 arcminute radius for stacked galaxies convolved with a 2.1' beam at a range of redshifts. Black line at $z=1$ represents best-fit power law to observational ACT data from \cite{meinke_2023}.  \label{therme_model}}
\end{figure}

Figure \ref{therme_model} shows the thermal energy versus stellar mass at each redshift. At the two higher redshifts, we find little difference between AGN and N-AGN at low stellar masses, with slightly more of a distinction at larger masses. At low redshifts, there is even more scatter in the results, and we do not find a strong trend between stellar mass and thermal energy. 

The third panel of Figure \ref{therme_model} includes ACT data and a power law fit to data in the central mass bins from \cite{meinke_2023}. Here we see that the observational data agrees well with our AGN sample, although the power law has a steeper relationship than seen in SIMBA. It is not clear if a power law is truly the best model for this data, but as there are large uncertainties in stellar mass in the high and low observational bins, fitting to the central and most populated bin is done in \cite{meinke_2023}. The flat slope in the stellar mass-thermal energy relationship relative to the forward-modeled power law was thought to be due to observational uncertainties, but we see a similarly flattened slope in SIMBA despite knowing the stellar masses with certainty. A large part of this result is most likely due to the low sample sizes at high masses (see Figure \ref{galsamp}), so it is difficult to discern a relationship, nor do we see any clear differences between AGN and N-AGN.

\begin{figure}[t]
\includegraphics[width = \linewidth]{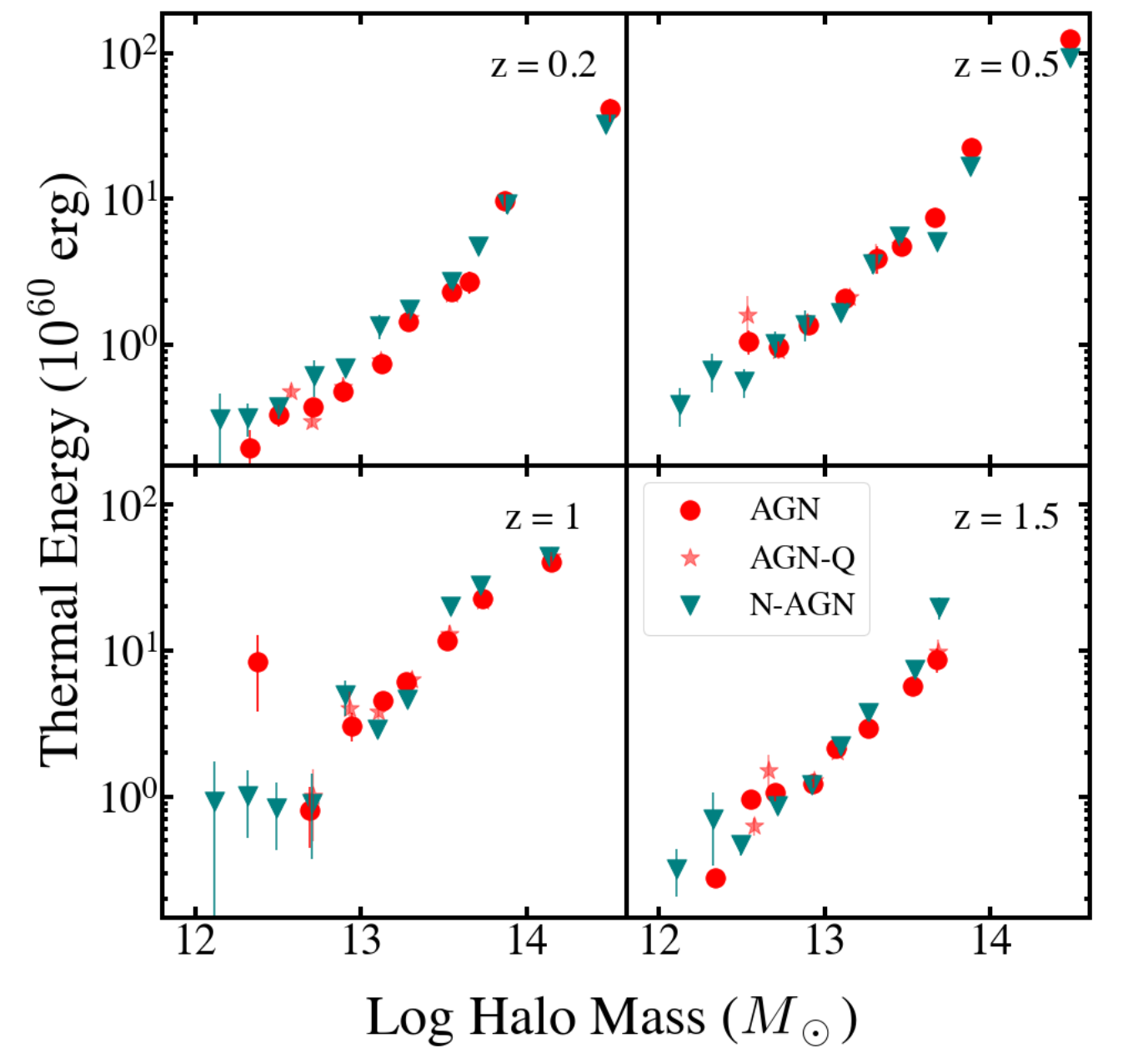}
\caption{Thermal energy measured within a 2 arcminute radius for massive galaxies convolved with a 2.1' beam plotted against halo mass at a range of redshifts.  \label{therme_halo}}
\end{figure}

By contrast, we do find a stronger correlation between halo mass and thermal energy, as seen in Figure \ref{therme_halo}. Here, we have once again stacked on galaxies, but we present the results binned by the mass of the halo each galaxy belongs to. The scatter and uncertainties are significantly reduced when plotting against halo mass, although there is almost no difference between the AGN and N-AGN runs. As opposed to the stellar mass results, we find that N-AGN actually has a slightly larger thermal energy than AGN at high halo masses. 

The difference in the results from stellar versus halo mass stacks indicates that the primary impact of AGN feedback in SIMBA is to change the stellar-mass halo-mass relationship. This is likely due to the fact that the jet-dominated feedback in SIMBA is largely kinetic, with no strong impact on the temperature of the gas as a function of halo mass. AGN feedback simply causes galaxies at a given stellar mass to be hosted in larger-mass halos, and thus stacking by stellar mass leads to stronger signals primarily because of the increase in gravitational heating from these halos.  On the other hand, the dependence of the tSZ signal with halo mass is likely to be different in simulations that adopt more bursty ``quasar mode" interactions  \citep[e.g.][]{Scannapieco_2004} as the primary source of quenching. This suggests that using a combination of stacks by halo mass and stacks by stellar mass may prove to be a fruitful method for using future tSZ data sets to constrain AGN feedback.

\section{Conclusions}\label{sec5}

AGN feedback is the leading explanation of the observed cosmic downsizing trend, whereby star formation is quenched in massive galaxies starting at $z\approx2$. The heating done by AGN is likely to leave measurable signatures in the CGM, which can be probed a millimeter wavelengths using the redshift-independent thermal Sunyaev Zel'dovich effect. Here, we used the SIMBA cosmological galaxy formation simulations to quantify these signatures and their dependence on AGN feedback. 

We used a stacking approach to model the tSZ profile in a sample of galaxies with $M_\text{stellar}>10^{11}M_\odot$, and convolved the data with two separate beams. The first, ACT's 2.1' beam, was used to compare against recent $z \approx 1$ results from \cite{meinke_2023}. We found that AGN feedback leads to a slightly higher signal at small radii, and observational data aligns well with SIMBA's predictions. We also compared against ACT data at $z \approx 0.5$ \citep{Schaan_2021}, finding a much weaker agreement between SIMBA and observation, with a significantly stronger signal predicted by SIMBA's AGN run. This suggests that SIMBA's treatment of AGN feedback may inject too much energy at late times. 

We also generated radial profiles at ACT resolution at a redshift of 0.2 and 1.5, to provide $y$-profiles that could be used in future work to better understand SIMBA's treatment of AGN at a range of cosmic times. We found the greatest difference between AGN and N-AGN at $z \approx 0.5$, suggesting further observational studies at that redshift and at the resolutions achievable by ACT or SPT could be beneficial both to confirm detections of AGN feedback as well as refine theoretical models.

We also convolved our data with the 9.5'' beam of the new TolTEC camera in order to provide a standard to compare against upcoming observations. With the higher-resolution beam, lower redshifts once again show the largest differences. Convolving with a higher-resolution beam also increased the overall signal of the stacked galaxies, suggesting detection may be easier with higher-resolution instruments. 

Most importantly, the absence of AGN leads to a more peaked profile within the virial radius, as probed by a high-resolution beam. This behavior is not seen with a larger beam, but it provides crucial evidence of how jet-dominated AGN feedback of the type implemented in SIMBA impacts the CGM distribution and evolution. SIMBA's feedback operates as a combination of velocity kicks and heating, so warm gaseous material will be pushed to larger radii, as is particularly apparent at redshifts of 0.5 and 1. As the gas is expelled to larger volumes while maintaining the same temperature, its density decreases, making cooling more difficult. 

New in this work is the use of moments to extract information about galaxy asymmetries without needing to know the orientation of the galaxies prior to stacking. The process and results outlined in \S \ref{moments} demonstrate that there is a large asymmetrical component to the galaxy's signal, most likely due to the presence of an asymmetric structure around these galaxies. By a redshift of 0.5, this asymmetry is more pronounced.  This is most likely due to galaxies clustering together over time,  with a significant number of large galaxies being found near the edges of massive halos, leading to a dipole in the signal around them.  Regardless of redshift, we found that the asymmetries are more pronounced for the SIMBA run without AGN, which can be attributed to AGN feedback's effective smoothing of the gas distribution. Stacking these transformed maps of the tSZ signal could provide a means of characterizing the distribution of galaxies beyond the information gained by measures such as the two-point correlation function. 

In addition to generating radial profiles, we also looked at the thermal energy measurements within a radius corresponding to roughly 2 Mpc-cm. When comparing the thermal energy for galaxies binned by stellar mass, we find a higher thermal energy in the run with AGN at large stellar masses, although there is significant scatter in the data, and the sample sizes are too small to draw any definitive conclusions. 

By contrast, binning galaxies by halo mass, we notably find no difference in the thermal energy measurements between the AGN and N-AGN run.  This can be reconciled with the stellar mass thermal energy measurements if the primary impact of AGN feedback in SIMBA is to change the stellar-mass halo-mass relationship rather than heat the gas.  This is also consistent with the fact that the jet-dominated treatment in SIMBA is largely kinetic rather than thermal. AGN feedback simply causes galaxies at a given stellar mass to be hosted in larger-mass halos, and thus stacking by stellar mass leads to stronger signals primarily because of the increase in gravitational heating rather than feedback heating. 

This goes against expectations from studies in which AGN feedback is modeled as dominated by more bursty ``quasar mode" interactions  \citep[i.e.][]{Scannapieco_2004}, which predict feedback to almost double the thermal energy compared to purely gravitational heating \citep{Spacek_2016}.  Thus the unexpected lack of a clear difference in these results is most likely due to AGN feedback being treated as a softer kick, as opposed to an explosive mechanism. This suggests that a combination of stacks by halo mass and stacks by stellar mass may prove to be an effective approach for applying thermal energy measurements to distinguish between modes of AGN feedback.

With the development of better instruments, the tSZ signal is proving a useful tool in distinguishing feedback models at high redshift. The success of SIMBA in replicating $z\approx 1$ results from ACT shows promise for future work using a comparison of simulations and observations to pin down AGN feedback behavior from upcoming projects such as TolTEC. 

\section*{Acknowledgements}

We would like to acknowledge Peter Behroozi, Rafael Gavazzi, Neal Katz, Philip Mauskopf, Jeremy Meinke, Elena Pierpaoli, and Lisiyuan Yang for helpful discussions, as well as the anonymous referee for helpful comments that greatly improved this manuscript.   We also thank Robert Thompson for developing CAESAR, and the yt team for the development and support of this useful community tool.  SG and ES acknowledge support from NASA grant 80NSSC22K1265.  We would also like to thank the Kavli Institute for Theoretical Physics and the organizers of the workshop The Cosmic Web: Connecting Galaxies to Cosmology at High and Low Redshift (CosmicWeb23). This research was supported in part by the National Science Foundation under Grant No. NSF PHY-1748958.

\appendix

\section{Correlation Matrices}\label{appa}

\begin{figure*}[ht!]

\centering
{\includegraphics[width=0.45\textwidth]{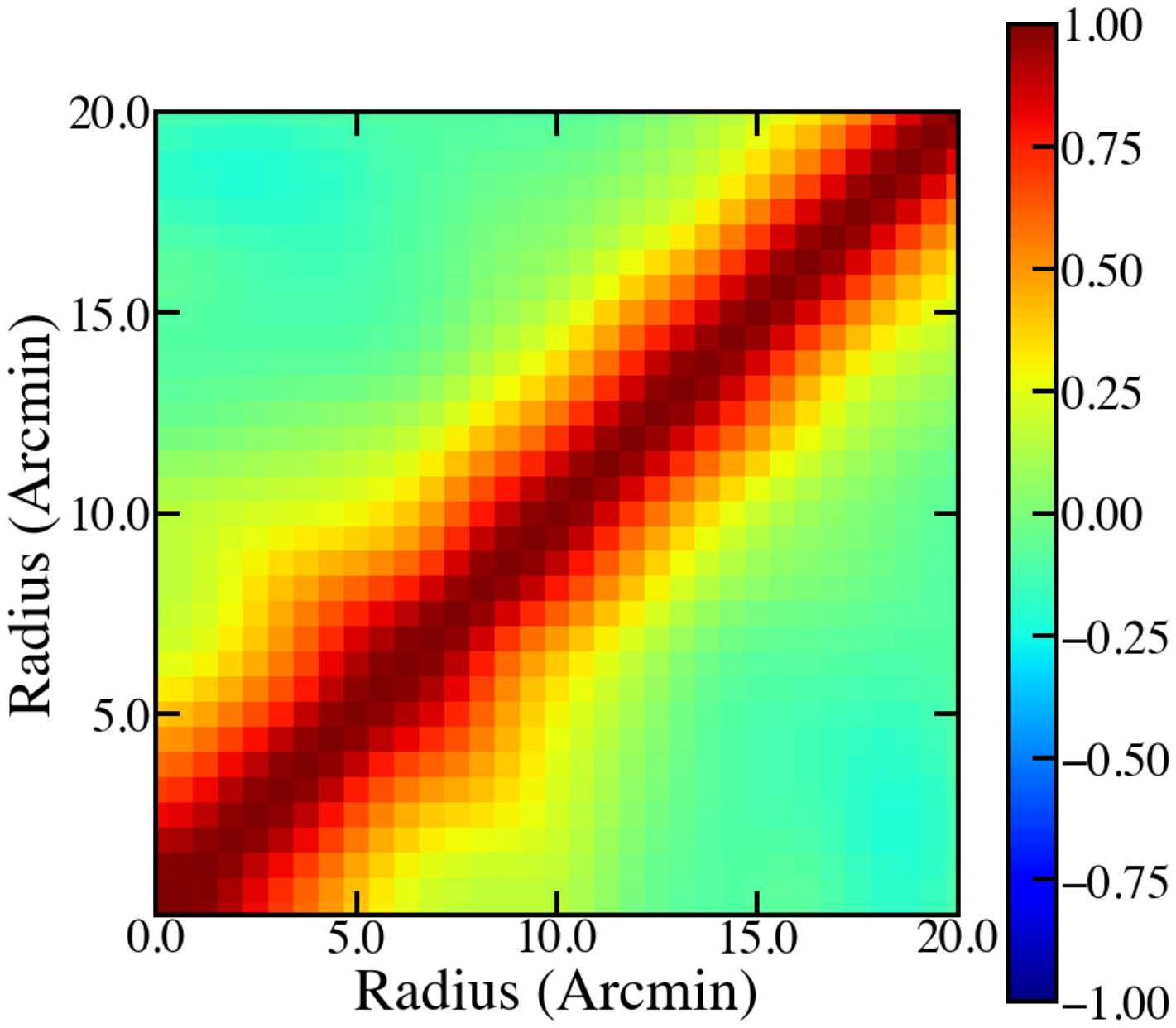}}
{\includegraphics[width=0.45\textwidth]{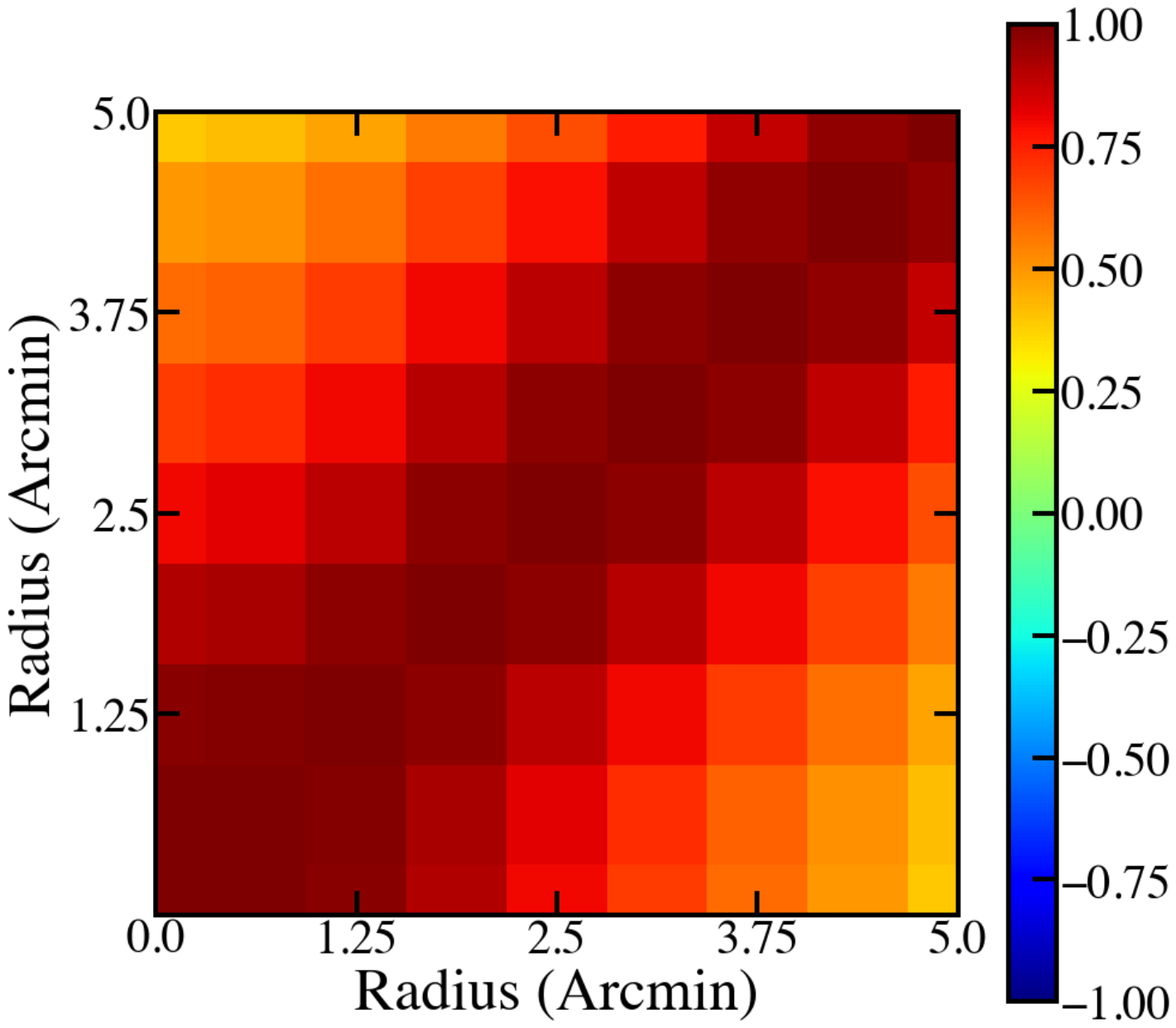}}
\caption{Radial bin correlation matrices for 37.5'' bins for SZ-y signals convolved with the ACT 2.1' beam at $z=1.$ Left: Correlation matrix out to radial distances of 20 arcminutes for the stacked AGN data, Right: Correlation matrix out to a radius of 5 arcminutes. The radial profiles shown in Figure \ref{actcomp} represent bins of this size. \label{spt_corr}}

\end{figure*}

\begin{figure*}[ht!]

\centering
{\includegraphics[width=0.45\textwidth]{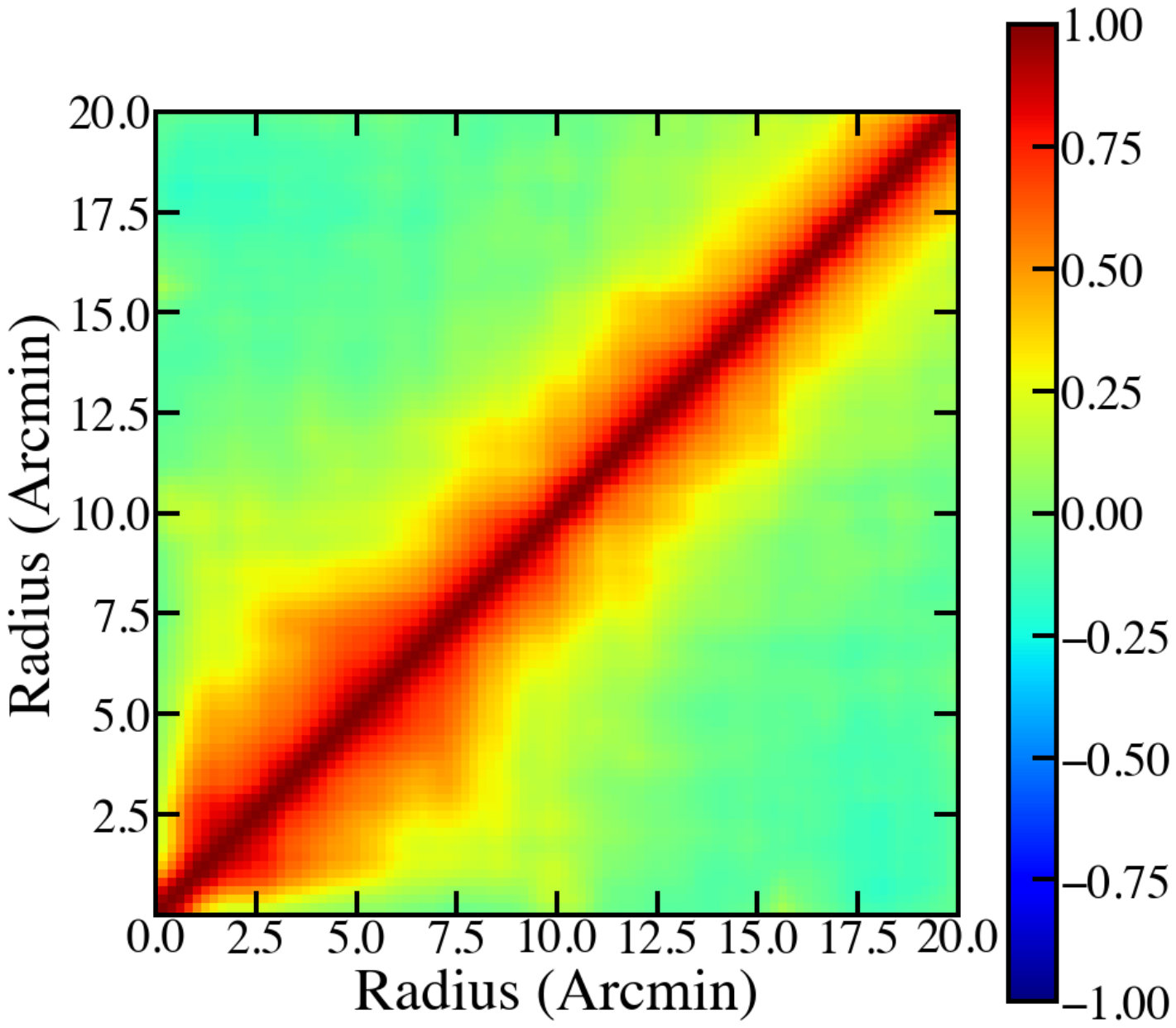}}
{\includegraphics[width=0.45\textwidth]{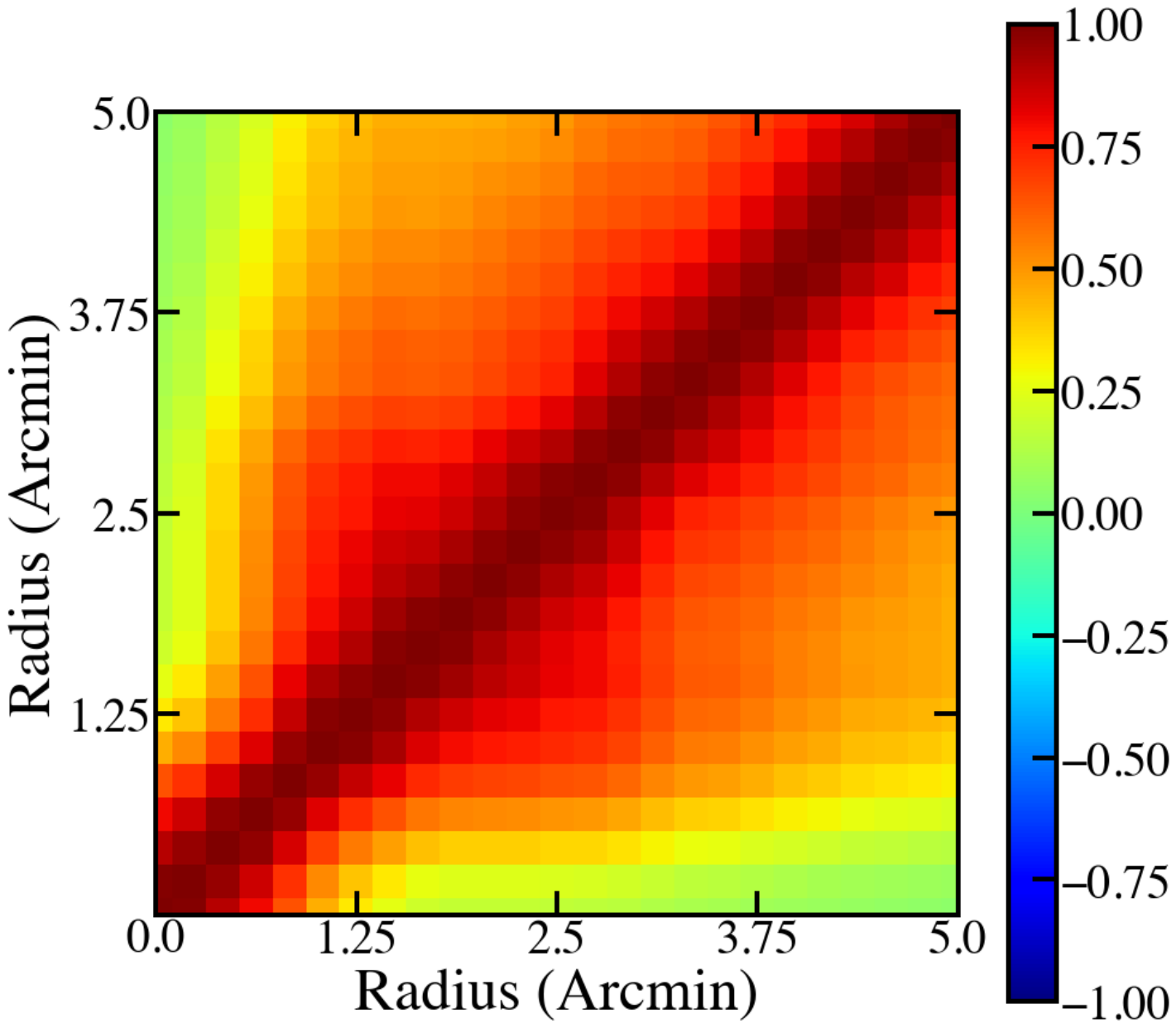}}
\caption{Radial bin correlation matrices for 12.5'' bins for SZ-y signals convolved with TolTEC's 9.5'' beam at $z=1$. Left: Correlation matrix out to radial distances of 20 arcminutes for the stacked AGN data, Right: Correlation matrix out to a radius of 5 arcminutes. \label{toltec_corr}}

\end{figure*}

We set the radial bins to reduce the correlation between neighboring pixels, as shown in Figures \ref{spt_corr} and \ref{toltec_corr}. We show only the bins for a redshift of one, but a similar approach was taken with each of the four redshifts. As the pixel size of our maps is 2.5'', we never have radial bins with less than 4 pixels (10''), which would correspond to a higher-resolution than TolTEC's 9.5'' beam. The bin size at each redshift for the two beams is summarized in Table \ref{bins}. While the radial bins are not entirely independent, this method of altering the pixel size of each bin significantly improves the correlation between neighboring points in the radial profiles.

\begin{table}[h!]
    \centering
    \caption{Bin Sizes for Radial Profiles}
    \begin{tabular}{ c c c }
    \hline
    \hline
         Redshift & ACT Bins & TolTEC Bins  \\
         \hline
         0.5 & 50''  &  25'' \\
         1 & 37.5'' &  12.5'' \\ 
         1.5 & 25'' & 10'' \\ 
         2 & 25'' & 10''\\ 
         
         \hline
    \end{tabular}

    \label{bins}
\end{table}

\vspace{5mm}

\newpage

\bibliographystyle{aasjournal}
\bibliography{bib.bib}

\end{document}